\documentclass{article}
\usepackage{arxiv}

\usepackage[caption = false]{subfig}

\usepackage[utf8]{inputenc} 
\usepackage[T1]{fontenc}    
\usepackage{hyperref}       
\usepackage{url}            
\usepackage{booktabs}       
\usepackage{amsfonts}       
\usepackage{nicefrac}       
\usepackage{microtype}      
\usepackage{lipsum}
\usepackage{amsmath,mathtools,bbm,multirow,url,hyperref,enumerate}
\usepackage{color}
\usepackage{float}
\usepackage{hyperref}
\usepackage{makecell}
\usepackage{algorithm,algorithmic}
\usepackage{multicol}
\usepackage{natbib}
\usepackage{hyperref}
\usepackage{soul}
\hypersetup{colorlinks,citecolor=blue}
\title{Modeling synergism in early phase cancer trials with drug combination with continuous dose levels: is there an added value?}

\author{
\normalsize{Mourad Tighiouart$^1$, Jos\'e L. Jim\'enez$^2$, Marcio A. Diniz$^1$, Andr\'e Rogatko$^1$}\\
$^1$ Cedars-Sinai Medical Center, California, USA\\
$^2$ Novartis Pharma AG, Basel, Switzerland
}

\begin{document}
\maketitle
\begin{abstract}
In parametric Bayesian designs of early phase cancer clinical trials with drug combinations exploring a discrete set of partially ordered doses, several authors claimed that there is no added value in including an interaction term to model synergism between the two drugs. In this paper, we investigate these claims in the setting of continuous dose levels of the two agents. Parametric models will be used to describe the relationship between the doses of the two agents and the probability of dose limiting toxicity and efficacy. Trial design proceeds by treating cohorts of two patients simultaneously receiving different dose combinations and response adaptive randomization. We compare trial safety and efficiency of the estimated maximum tolerated dose (MTD) curve between models that include an interaction term with models without the synergism parameter with extensive simulations. Under a selected class of dose-toxicity models and dose escalation algorithm, we found that not including an interaction term in the model can compromise the safety of the trial and reduce the pointwise reliability of the estimated MTD curve.
\end{abstract}


\section{Introduction}


The primary goal of cancer phase I trials is to study the safety of new cytotoxic, biologic, or immunotherapy drugs or combination of existing agents with possibly radiation therapy and to determine the maximum tolerated dose (MTD) for use in future efficacy studies. These trials enroll advanced stage cancer patients who have exhausted conventional treatment (\cite{Roberts2004}). For ethical considerations and to minimize the number of subjects exposed to toxic doses, patients are accrued to these trials sequentially and dose allocation is adaptive and depends on the doses given to current and previously treated patients and their toxicity outcomes. For treatments where only one drug is allowed to vary during the trial, and assuming a non-decreasing relationship between the dose levels and the probability of dose limiting toxicity (DLT), the MTD is defined as the maximum dose $\gamma$ that will produce DLT in a prespecified proportion $\theta$ of patients (\cite{Gatsonis1992}) 

\begin{equation}
\label{defmtd}
P(DLT |{\rm dose}=\gamma)=\theta.
\end{equation}

Typically, DLT is a binary indicator of toxicity and is defined as a grade 3 or 4 non-hemathologic or grade 4 hematologic toxicity, although models with quasi-continuous (\cite{Chen2010,Chen2012}) and ordinal toxicity grades (\cite{Dressler2011,Iasonos2011,Tighiouart2012ordinal,diniz2020bayesian}) have been studied for single agent and drug combination dose-finding trials. Statistical designs for dose-finding studies have been extensively studied over the last three decades and include model-based designs such as the continual reassessment method (CRM) (\cite{o1990continual}) and its modification (\cite{goodman1995some,korn1991selecting,moller1995extension,o1996continual,leung2002extension,o2003continual,Iasonos2011,daimon2011posterior,liu2013bayesian}), escalation with overdose control (EWOC) (\cite{babb1998cancer}) and its variants and extensions (\cite{Tighiouart2005flexible,tighiouart2010dose,Chen2012,Tighiouart2012,Tighiouart2012covariate,tighiouart2014dose,wheeler2017toxicity,tighiouart2018bayesian}), semi-parameteric design (\cite{clertant2017semiparametric}) and nonparametric dose-finding methods, e.g., the modified toxicity probability interval (\cite{ji2013modified}), the Bayesian optimal design (\cite{yuan2016bayesian}), and the nonparametric overdose control method (\cite{lin2017bayesian}).

To reduce treatment resistance to therapy, a combination of cytotoxic, biologic, and immunotherapy drugs is often used in order to target different signaling pathways simultaneously and hence achieve higher response rates by using additive or synergistic agents. When all these drugs are allowed to vary during the trial, a challenging problem is to identify a subset of
these combinations among a larger set of permissible doses that will produce the same DLT rate $\theta$ since the MTD is no longer unique. Model-based designs for drug combinations that recommend a single combination for use in future efficacy studies have been proposed by \cite{yin2009bayesian,yin2009Bayesian1,wages2011continual,Shi2013,Riviere2014} among others. However, since different MTDs may have different levels of efficacy, we argue for the use of drug combination designs that recommend more than one MTD and study their efficacy in future randomized or response adaptive designs. To this end, and for discrete dose levels, \cite{thall2003dose} proposed a six-parameter dose-toxicity model and a two-stage algorithm that recommends three MTDs at the conclusion of the trial. Other parameteric and nonparametric designs that can recommend more than one MTD include the three-parameter regression model of \cite{wang2005two}, a Bayesian hierarchical model by \cite{Braun2010}, the product of independent beta probabilities (PIPE) \cite{mander2015product}, the two-dimensional Bayesian optimal interval design (2dBOIN) (\cite{lin2017nonparametric}), and the fully nonparametric approach of \cite{Razaee2022}. In cancer treatment using novel agents, the use of continuous dose levels is very common, particularly when the drugs are delivered as infusions intravenously. \cite{tighiouart2014dose} introduced a dose-finding method for continuous dose levels based on conditional EWOC and recommended an MTD curve at the conclusion of the trial. The method was extended to three drugs (\cite{Tighiouart2016}), case where the MTD curve lies anywhere in the Cartesian plane (\cite{tighiouart2017bayesian}), using the CRM (\cite{2017bayesiandiniz}), adjusting for a baseline covariate (\cite{diniz2018bayesian}), settings where an unknown fraction of DLTs is attributable to one or more agents (\cite{jimenez2019cancer}), using an ordinal toxicity grade \cite{diniz2020bayesian}, and to phase I/II designs (\cite{tighiouart2019two,jimenez2020bayesian, jimenez2021combining,jimenez2021bayesian}). 

In the above model-based designs for drug combinations, several authors argued for the unnecessary use of an interaction term in the model due to the inherent small sample size of these trials and inability to estimate the synergistic parameter of interest with good precision (\cite{wang2005two,Riviere2014,lyu2019aaa}). In fact, \cite{Mozgunov2021} carried out simulation studies to assess the effect of omitting the interaction term on the percent of dose recommendation in the logistic model of \cite{Riviere2014}. In general, they found that only marginal benefits are seen when using an interaction term and in the absence of prior information regarding the extent of synergism between the two drugs, there is no added benefit in including the interaction coefficient. In this manuscript, we investigate these claims in the setting of drug combinations with continuous dose levels using the design of \cite{tighiouart2017bayesian, Tighiouart2022}. We show using simulation studies that when the two drugs are highly synergistic, safety of the trial is compromised when omitting the interaction term from the model and efficiency of the estimated MTD curve as assessed by the pointwise bias and percent selection is greatly reduced.

The article is organized as follows. In Section 2, we describe the dose-toxicity model for drug combinations and review the dose-escalation algorithm based on the conditional EWOC scheme. Simulation studies comparing the performance of models with and without an interaction term are found in Section 3 and a discussion of these findings is included in Section 4.

\section{Phase I Design}
\label{sc_phase_1}

We consider the setting of two drugs $A$ and $B$ with continuous dose levels $x$ and $y$ in the intervals $[X_{\rm min}, X_{\rm max}]$ and $[Y_{\rm min}, Y_{\rm max}]$, respectively. We further standardize the doses to be in the interval $[0, 1]$ using the transformations $h_1(x)=(x-X_{\rm min})/(X_{\rm max} - X_{\rm min})$ and $h_2(y)=(y-Y_{\rm min})/(Y_{\rm max} - Y_{\rm min})$. 

\subsection{Dose-Toxicity Model}

We assume a class of dose-toxicity models for dose allocation and MTD curve estimation of the form
\begin{equation}
\label{toxmodel}
P(T=1 |x,y)=F(\beta_{0}+\beta_{1}x+\beta_{2}y+\beta_{3}xy),
\end{equation}where $T$ is the indicator of DLT, $T = 1$ if a patient given the dose combination $(x,y)$ exhibits DLT,  and $T = 0$ otherwise, the interaction parameter $\beta_{3} \geq 0$, and $F$ is a known cumulative distribution function (cdf). Finally, we  assume that that the probability of DLT is non-decreasing with the dose of any one of the agents when the other one is held constant. A necessary and sufficient condition for this property to hold is to assume $\beta_{1}, \beta_{2} > 0$. This is a common assumption for cytotoxic, biologic, immunotherapy drugs or radiation treatment. By definition, the MTD is any dose combination $(x^{*}, y^{*})$ such that $P(T=1|x^{*}, y^{*})=\theta$. The target probability of DLT $\theta$ is pre-specified by the clinician and depends on nature and severity of expected DLTs.

A convenient reparameterization of model (\ref{toxmodel}) in terms of parameters clinicians can interpret and to facilitate prior distribution elicitation is to use $\rho_{10}$, the probability of DLT when the levels of drugs $A$ and $B$ are 1 and 0, respectively, $\rho_{01}$, the probability of DLT when the levels of drugs $A$ and $B$ are 0 and 1, respectively, and $\rho_{00}$, the probability of DLT when the levels of drugs $A$ and $B$ are both 0. Under model assumption (\ref{toxmodel}), the MTD is the set of dose combinations

\begin{equation}
\label{defmtd1}
C =\left\{(x^{*},y^{*}): y^{*} = \frac{(F^{-1}(\theta)-F^{-1}(\rho_{00}))-(F^{-1}(\rho_{10})-F^{-1}(\rho_{00}))x^{*}}{(F^{-1}(\rho_{01})-F^{-1}(\rho_{00}))+\beta_{3}x^{*}}\right\}.
\end{equation}

Let $\pi(\rho_{00},\rho_{01},\rho_{10},\eta_3)$ be a prior distribution of the model parameters and $D_{k} =\{(x_{i},y_{i},T_{i}),i=1,\ldots,k\}$ be the data after enrolling $k$ patients to the trial. The posterior distribution is

\begin{equation*}
\begin{split}
& \pi(\rho_{00},\rho_{01},\rho_{10},\beta_{3}|D_k) \propto \prod_{i=1}^{k} \left[G(\rho_{00},\rho_{01},\rho_{10},\beta_{3};x_{i},y_{i})\right]^{T_{i}} \left[1-G(\rho_{00},\rho_{01},\rho_{10},\beta_{3};x_{i},y_{i})\right]^{1-T_{i}} \\ 
& \times \pi(\rho_{00},\rho_{01},\rho_{10},\eta_3),   
\end{split}
\end{equation*}where 

\begin{equation}
\label{Gfunction}
\begin{split}
& G(\rho_{00},\rho_{01},\rho_{10},\beta_{3};x_{i},y_{i})=F\left[F^{-1}(\rho_{00})+(F^{-1}(\rho_{10})-F^{-1}(\rho_{00}))x_{i} \nonumber \right. \\
& + (F^{-1}(\rho_{01})-F^{-1}(\rho_{00}))y_{i}+\beta_{3}x_{i}y_{i}
\end{split}
\end{equation}

\subsection{Design}
We will use the design proposed by \cite{tighiouart2017bayesian} and applied to phase I/II trials by \cite{tighiouart2019two,jimenez2020bayesian} and to ordinal toxicity grades by \cite{diniz2020bayesian}. The design consists of treating successive cohorts of two patients where one subject receives a dose of drug $A$ for a given dose of drug $B$ that was previously assigned and the other subject receives a dose of drug $B$ for a given dose of drug $A$ that was previously allocated. This process allows for better exploration of the dose combination space. For safety considerations and to limit the number of patients exposed to toxic doses, we start the trial with the minimum dose combination and use the escalation with overdose control (EWOC) principle \cite{babb1998cancer,Tighiouart2005flexible,tighiouart2010dose,Tighiouart2012,wheeler2017toxicity}. We briefly review the algorithm:

\begin{enumerate}
    \item The first cohort of two patients receive the minimum dose combination 
    
    $(x_1,y_1)=(x_2,y_2)=(X_{\rm min},Y_{\rm min}).$ Let $D_2=\{(x_1,y_1,T_1),(x_2,y_2,T_2)\}.$
    \item In the $i-$th cohort of two patients,
    \begin{itemize}
        \item If $i$ is even, patients $(2i-1)$ and $2i$ receive dose combinations $(x_{2i-1},y_{2i-3})$ and $(x_{2i-2},y_{2i}),$ respectively, where
        
        $x_{2i-1}=\Pi_{\Gamma_{A|B=y_{2i-3}}}^{-1}(\alpha|D_{2i-2})$ and $y_{2i}=\Pi_{\Gamma_{B|A=x_{2i-2}}}^{-1}(\alpha|D_{2i-2})$.
        \item If $i$ is odd, patients $(2i-1)$ and $2i$ receive dose combinations $(x_{2i-3},y_{2i-1})$ and $(x_{2i},y_{2i-2}),$ respectively, where
        
        $x_{2i}=\Pi_{\Gamma_{A|B=y_{2i-2}}}^{-1}(\alpha|D_{2i-2})$ and $y_{2i-1}=\Pi_{\Gamma_{B|A=x_{2i-3}}}^{-1}(\alpha|D_{2i-2})$.
    \end{itemize}
    \item Repeat step 2 until a pre-specified number of patients $n$ are treated subject to a stopping rule for safety.
\end{enumerate}
   
Here, $\Pi_{\Gamma_{A|B=y}}$ is the posterior cumulative distribution function of the MTD of drug $A$ given that the dose of drug $B$ is $y$. This posterior distribution is easily obtained using the MCMC samples of the model parameters since $\Gamma_{A|B=y}$ and $\Gamma_{B|A=x}$ are functions of $\rho_{00},\rho_{01},\rho_{10}$, and $\eta_3$.

In step 2 of the above algorithm, a dose for drug $A$ is selected in such a way that the posterior probability of exceeding the MTD of drug $A$ given the current dose level of drug $B$ is bounded by a feasibility bound $\alpha$. For example, after resolving the DLT status of the first two patients, the third patient receives dose combination $(x_3,y_3)$ where $y_3 = y_1$ and $x_3$ is the $\alpha-$th percentile of $\Pi_{\Gamma_{A|B=y_1}}$, i.e., the largest dose such that the posterior probability of exceeding the MTD of $A$ given that $B=y_1$ is no more than $\alpha$. This is the defining property of EWOC.

{\it Stopping rule.} For ethical considerations, after $n_1$ patients are evaluable for toxicity, enrollment to the trial is suspended if there is statistical evidence that the minimum dose combination is too toxic, i.e., $P(P(T=1|(x,y)=(0,0)) > \theta +\xi_1 |D_{n_{1}}) > \xi_2$. Here, $\xi_1,
\xi_2$ are design parameters chosen to achieve good operating characteristics under various scenarios for the location of the true MTD curve. 

At the conclusion of the trial, an estimate of the MTD curve in (\ref{defmtd1}) is given by

\begin{equation}
\label{estmtd}
\hat{C} =\left\{(x^{*},y^{*}): y^{*} = \frac{(F^{-1}(\theta)-F^{-1}(\hat{\rho}_{00}))-(F^{-1}(\hat{\rho}_{10})-F^{-1}(\hat{\rho}_{00}))x^{*}}{(F^{-1}(\hat{\rho}_{01})-F^{-1}(\hat{\rho}_{00}))+\hat{\beta}_{3}x^{*}}\right\},
\end{equation}where $\hat{\rho}_{00}, \hat{\rho}_{01}, \hat{\rho}_{10},\hat{\beta}_3$ are the posterior medians given $D_n$.

\section{Simulation Studies}
In this section, we compare the performance of this design in estimating the MTD curve between the working model $F(\beta_{0}+\beta_{1}x+\beta_{2}y+\beta_{3}xy)$ and the working model that does not use an interaction term $F(\beta_{0}+\beta_{1}x+\beta_{2}y)$, with $F(\cdot)$ being the logistic function $F(u)=(1+e^{-u})^{-1}.$

\subsection{Scenarios and prior distributions}

\begin{figure}[h]
\centering
\subfloat[Scenario 1]{\includegraphics[width = 2.0in]{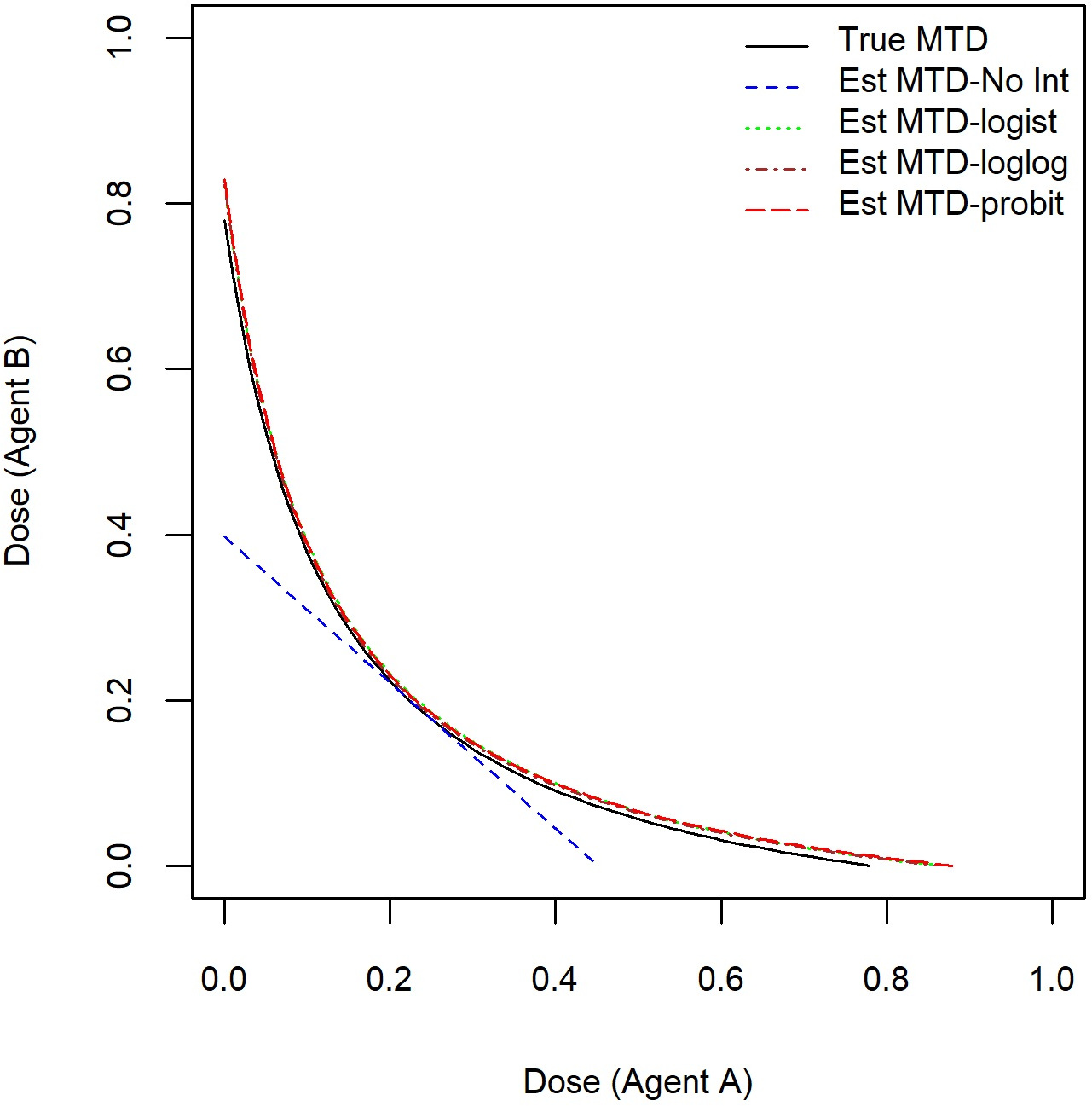}} 
\subfloat[Scenario 2]{\includegraphics[width = 2.0in]{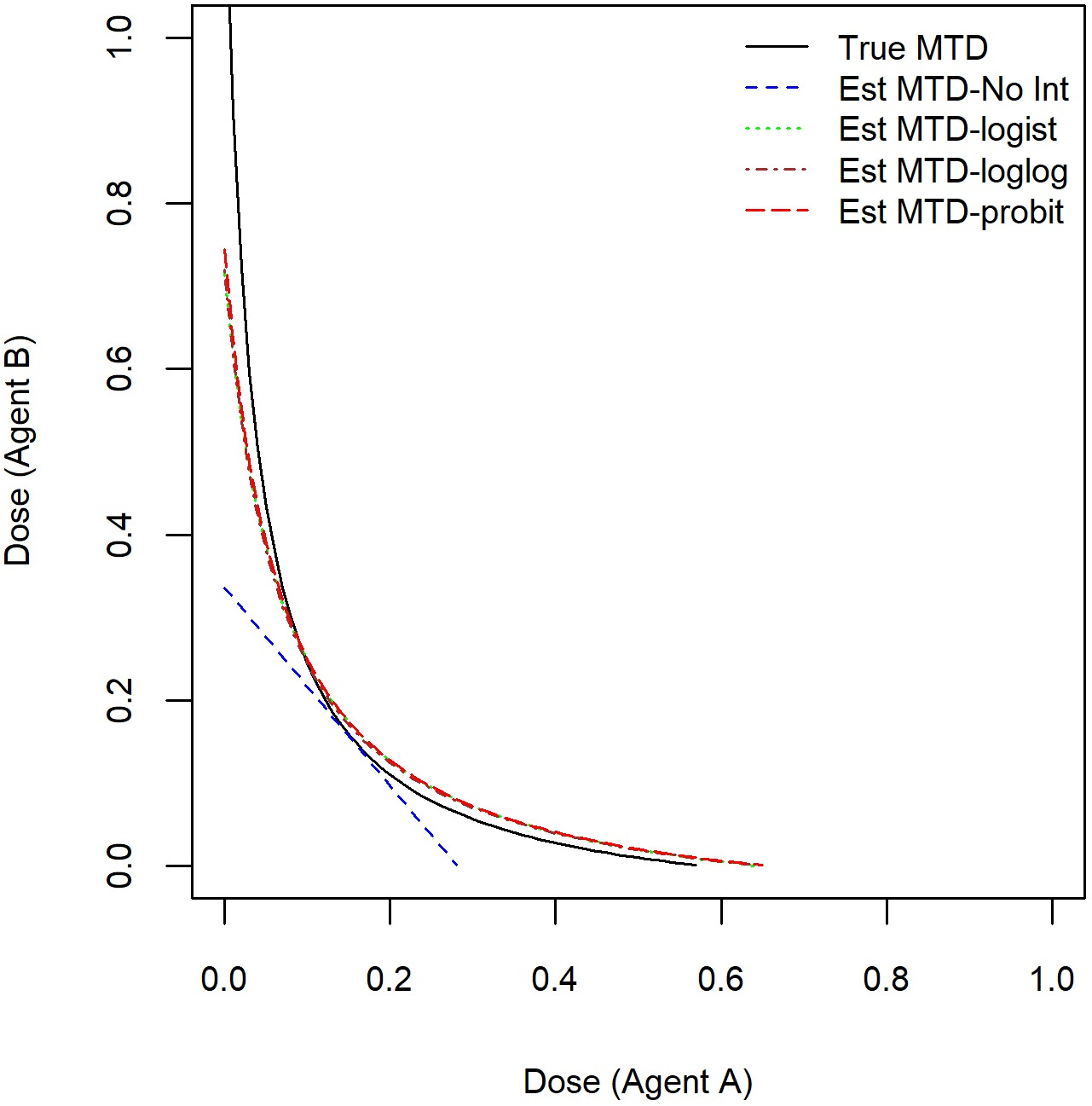}}\\
\subfloat[Scenario 3]{\includegraphics[width = 2.0in]{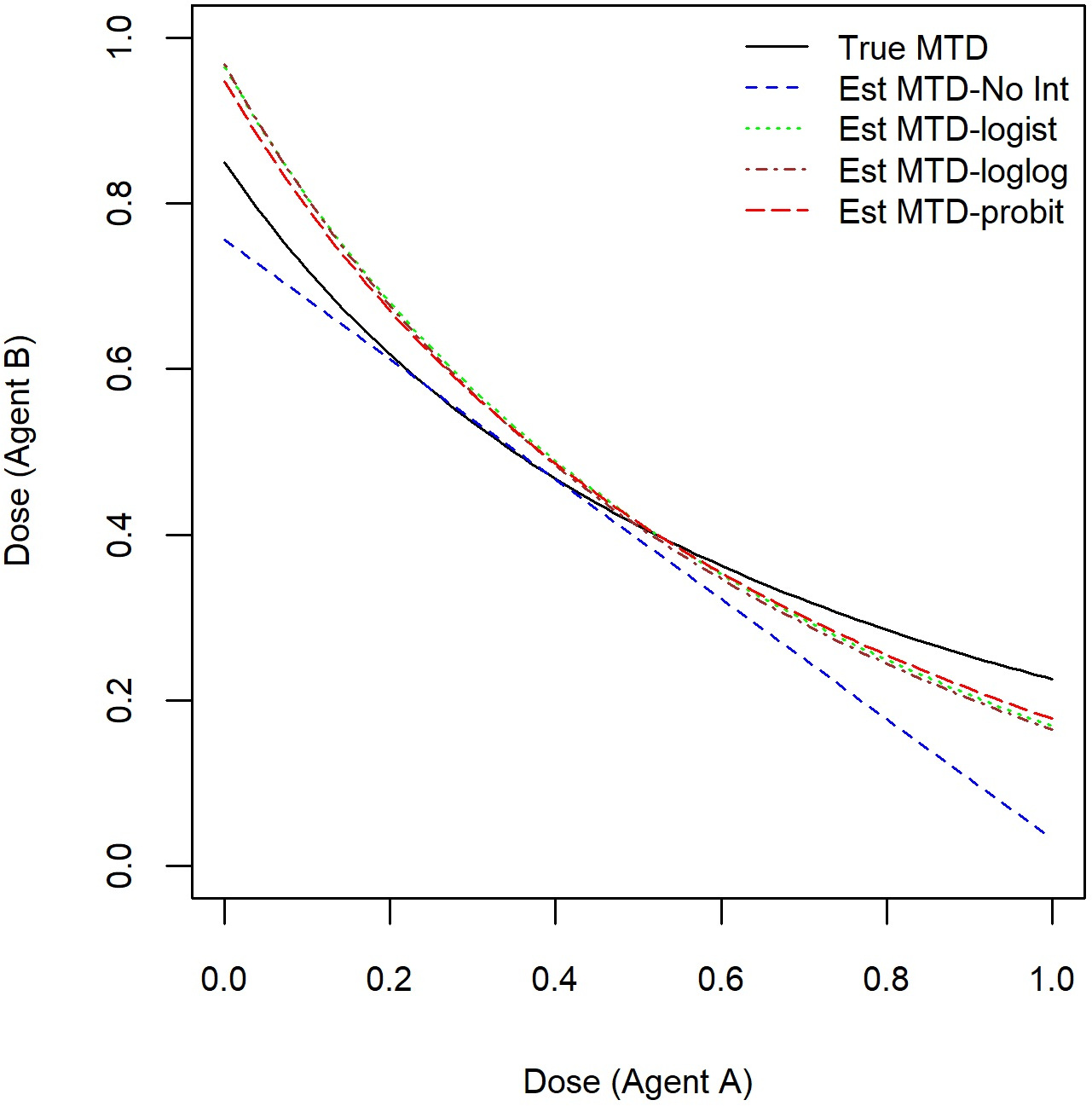}}
\subfloat[Scenario 4]{\includegraphics[width = 2.0in]{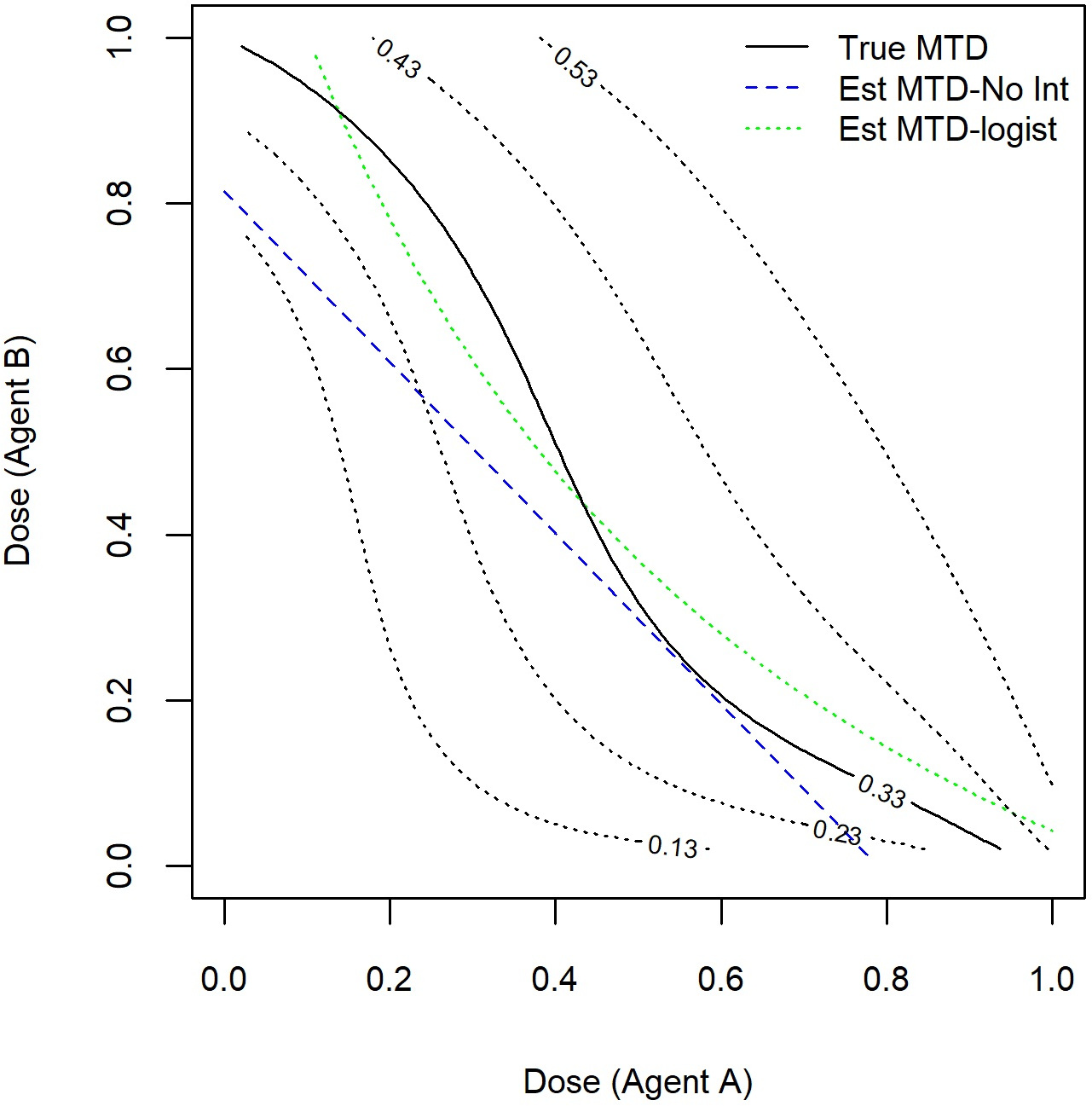}} 
\caption{True and estimated MTD curves from $m=2000$ trial replicates under scenarios 1--4.}
\label{fig:scenarios}
\end{figure}

We consider four scenarios for the location of the true MTD curve shown by a solid black curve in Figure~\ref{fig:scenarios}. For scenarios (1--3), estimates of the MTD curves are obtained using true and misspecified models for the dose-toxicity relationship. These are the logistic, probit $F(u)=\Phi(u),$ and complementary log-log $F(u)=1-e^{-e^{u}}$ link functions, where $\Phi(\cdot)$ is the cdf of the standard normal distribution. The misspecified models were chosen so that they have the same MTD curve as the logistic model by shifting the intercept coefficient. For example, when $F(\cdot)$ is the logistic function, points $(x^{\star},y^{\star})$ on the MTD curve satisfy $F^{-1}(\theta)= \beta_{0}+\beta_{1}x^{\star}+\beta_{2}y^{\star}+\beta_{3}x^{\star}y^{\star}$. The corresponding MTD curve for the probit model is any dose combination $(x^{\star},y^{\star})$ that satisfies $\Phi^{-1}(\theta)= \beta'_{0}+\beta'_{1}x^{\star}+\beta'_{2}y^{\star}+\beta'_{3}x^{\star}y^{\star}$. For the two models to have the same MTD curve, we set $\beta_i=\beta'_i, i=1, 2, 3$ and $\beta'_0=\beta_0+F^{-1}(\theta)-\Phi^{-1}(\theta).$ Figure~\ref{fig:contours}(a,b) show the extent of departure of the true logistic model from the models with probit and complementary log-log link functions under scenario 3. In scenario 4, a six-parameter model is used as the true model
$P(T=0|x,y)=(1+\alpha_{1}x^{ \beta_1}+\alpha_{2}y^{\beta_2}+\alpha_{3}(x^{\beta_1}y^{\beta_2})^{\beta_3})^{-1}.$ The true MTD curve and contour plots under this model are shown in Figure~\ref{fig:contours}(c). The true parameter values $(\rho_{00}, \rho_{00},\rho_{00},\beta_3)$ of the true logistic and six-parameter models are shown in the first column of Table~\ref{tab:safety}.

For each scenario, the target probability of DLT is $\theta=0.33,$ the trial sample size is $n=40$ patients, and $m=2000$ trial replicates were simulated to summarize the operating characteristics. A variable feasibility bound $\alpha$ was used starting with $\alpha=0.25$ and increasing this value in increments of 0.05 after each cohort of two patients are enrolled to the trial until $\alpha=0.5,$ see \cite{,chu2009,tighiouart2010dose}. For safety considerations, dose escalation of either drug during the trial is restricted to be no more than 20\% of the dose range of that drug.

The assumption $\beta_1, \beta_2 > 0$ implies that $0< \rho_{00} < {\rm min}(\rho_{01}, \rho_{10})$. Hence, we assumes that $\rho_{01}, \rho_{10}, \beta_3$ are independent {\it a priori}, $\rho_{01}, \rho_{10} \sim {\rm beta}(1,1),$ and conditional on $(\rho_{01}, \rho_{10})$, \\ $\rho_{00}/{\rm min}(\rho_{01}, \rho_{10}) \sim {\rm beta}(1,1)$. The prior on the interaction parameter $\beta_3$ is a gamma with mean 21 and variance 540, see \cite{tighiouart2014dose} for a justification of these hyperparameters.

\subsection{Operating characteristics}

Summary statistics for trial safety are reported as the percent of DLTs across all patients and all simulated trials and the percent of trials with DLT rates exceeding $\theta+0.05$ and $\theta+0.1$. We also report an estimate of the MTD curve

\begin{equation}
\label{estmtd1}
\bar{C} =\left\{(x^{*},y^{*}): y^{*} = \frac{(F^{-1}(\theta)-F^{-1}(\bar{\rho}_{00}))-(F^{-1}(\bar{\rho}_{10})-F^{-1}(\bar{\rho}_{00}))x^{*}}{(F^{-1}(\bar{\rho}_{01})-F^{-1}(\bar{\rho}_{00}))+\bar{\beta}_{3}x^{*}}\right\},
\end{equation}

\noindent where $F(\cdot)$ is the logistic function and $\bar{\rho}_{00}, \bar{\rho}_{01}, \bar{\rho}_{10}$ and $\bar{\beta}_{3}$ are the average posterior medians across all $m=2000$ simulated trials. 

\begin{figure}[h]
\centering
\subfloat[Contours of logistic and complementary log-log]{\includegraphics[width = 2.0in]{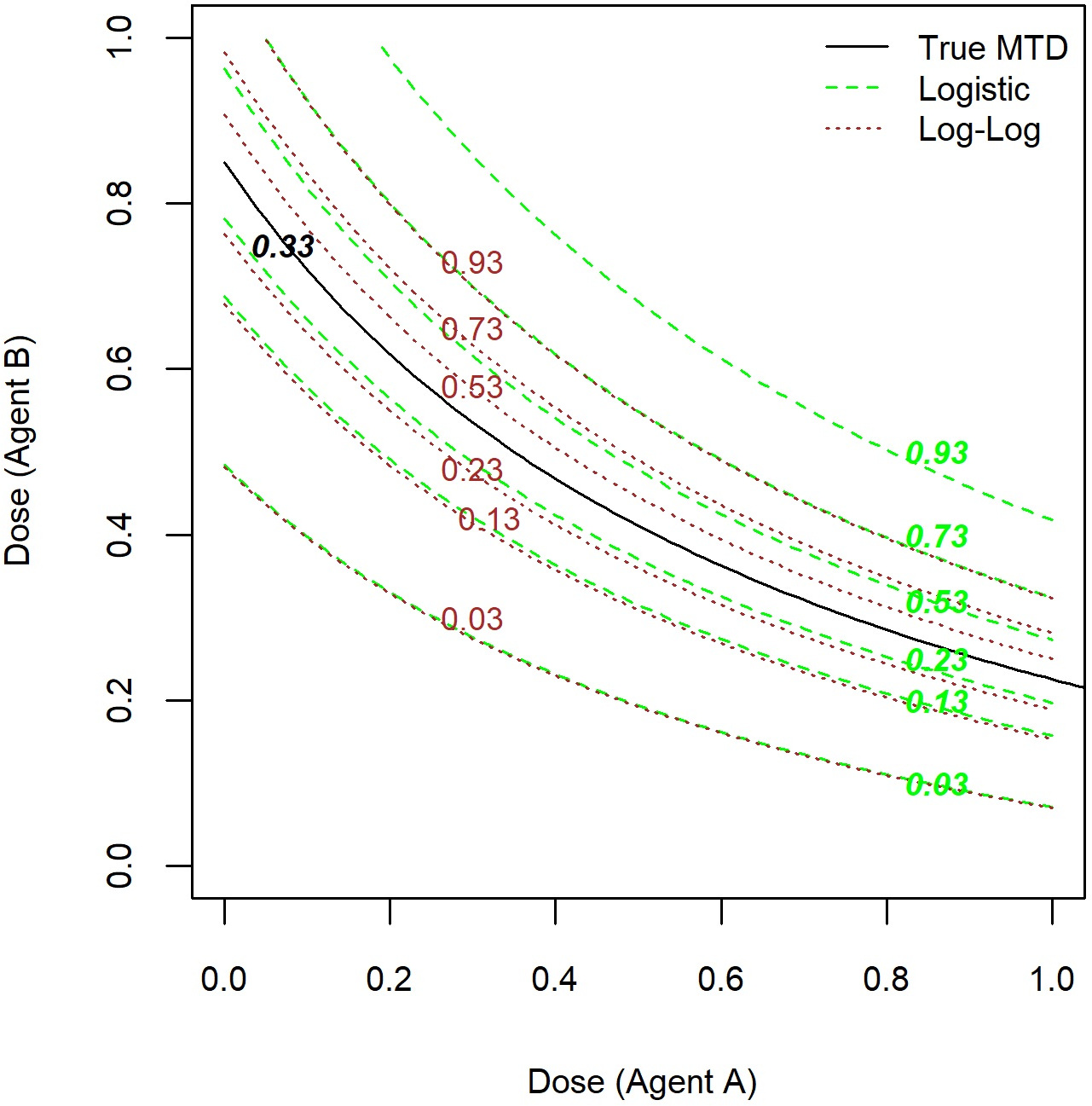}} 
\subfloat[Contours of logistic and probit]{\includegraphics[width = 2.0in]{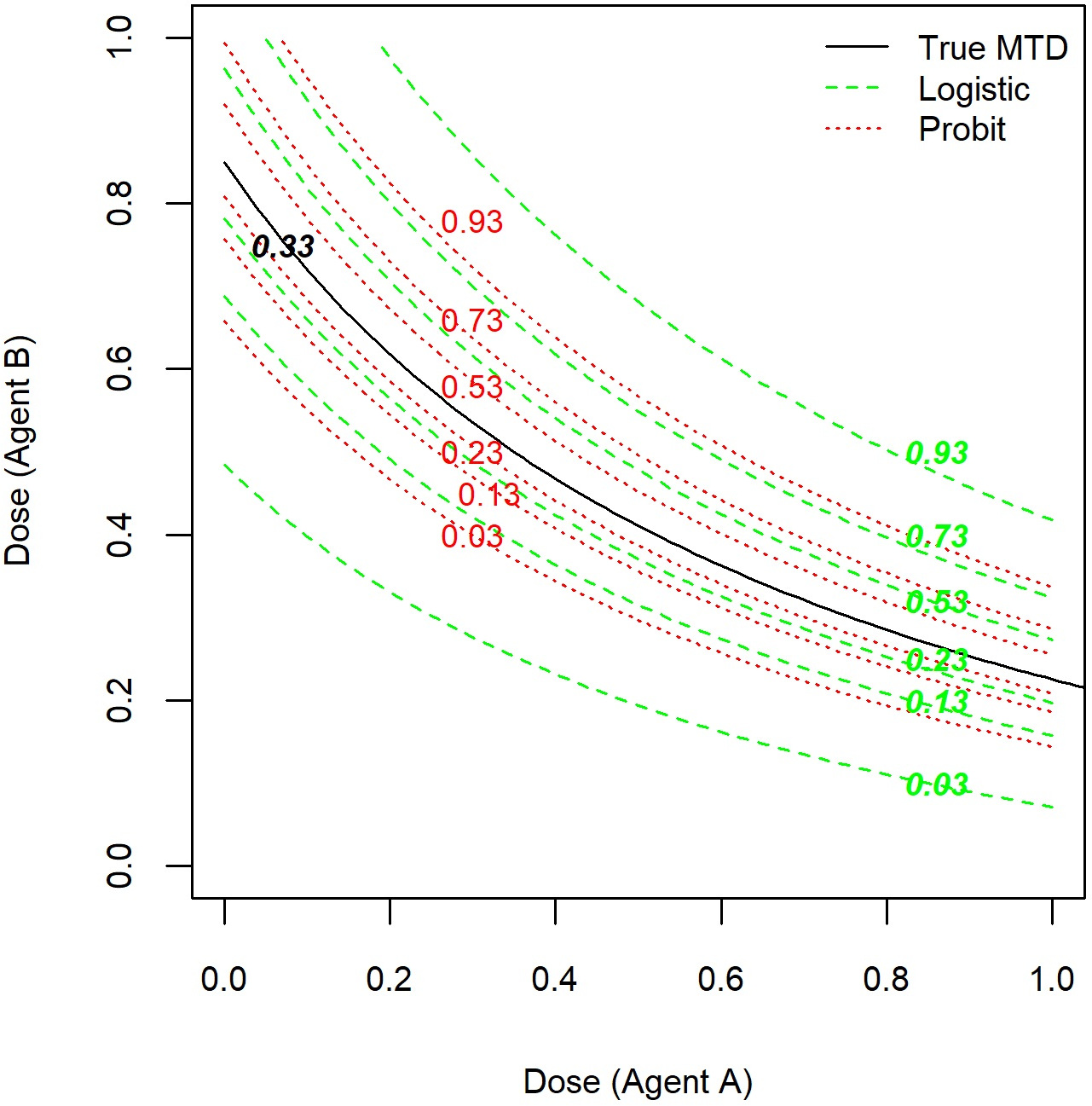}}\\
\subfloat[Contours of the six-parameter model]{\includegraphics[width = 2.0in]{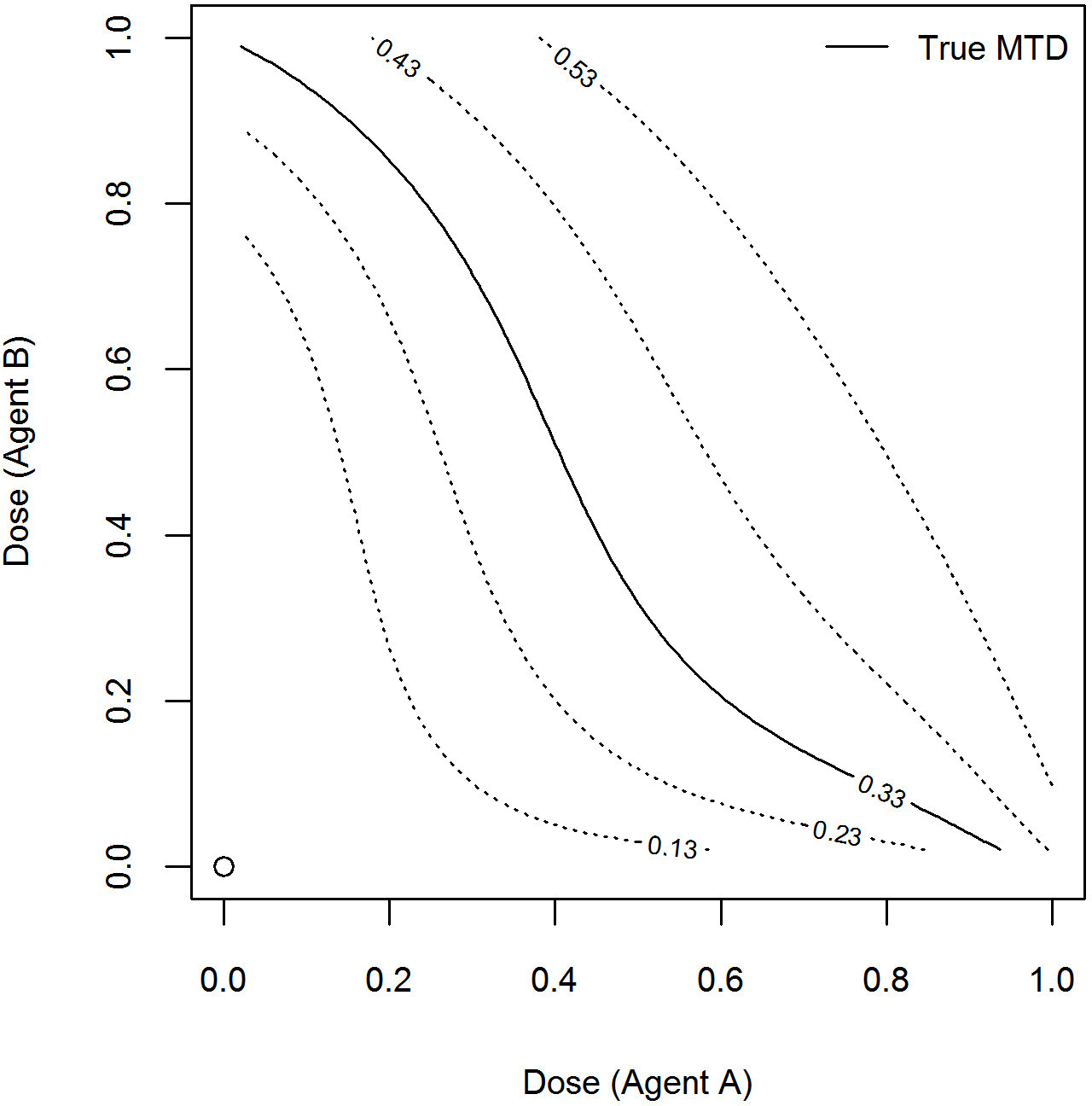}}
\subfloat[Selected neighborhoods for two MTD curves when $p=0.1$]{\includegraphics[width = 2.0in]{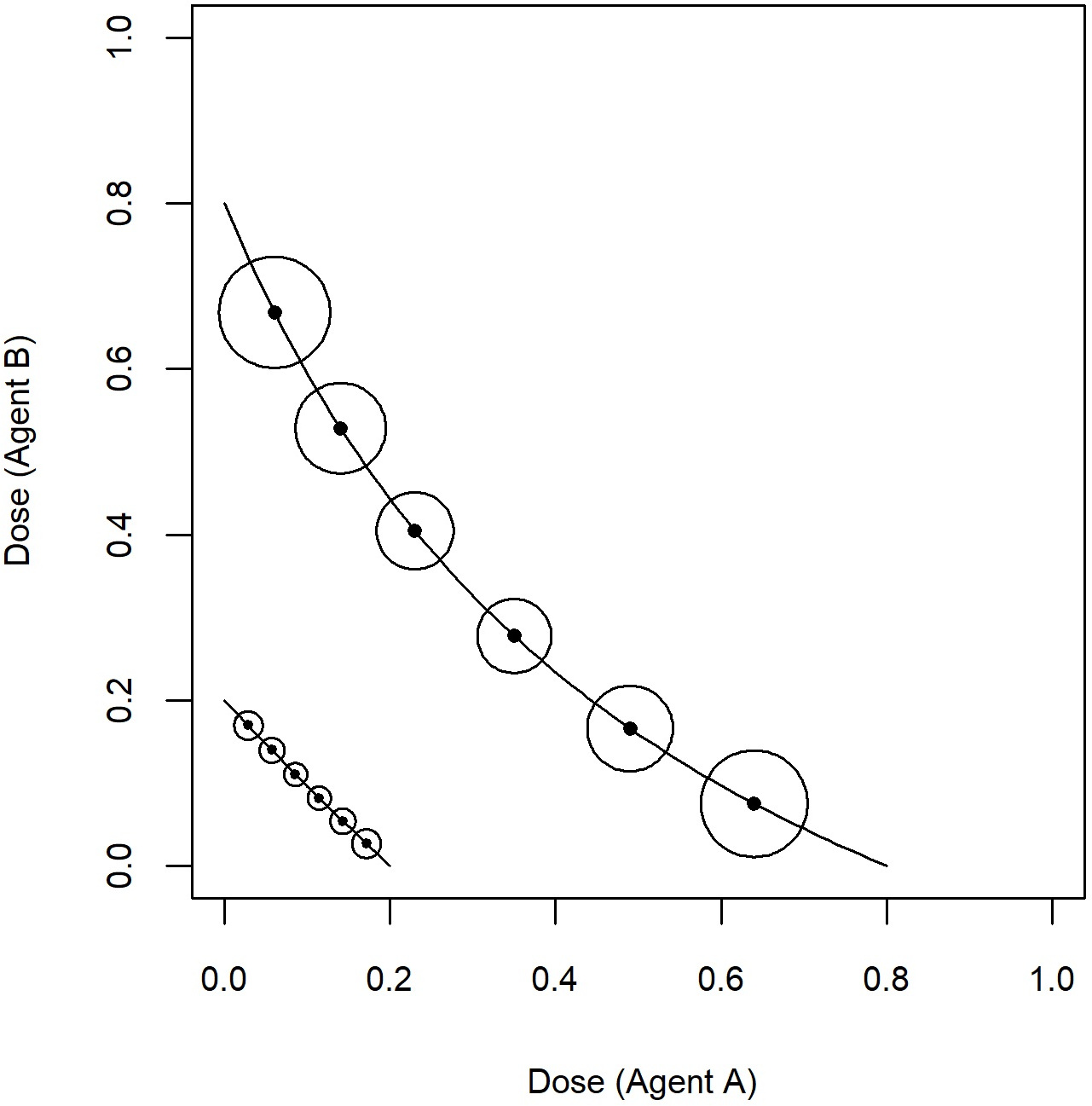}} 
\caption{Contour plots from the complementary log-log and probit models relative to the logistic model (a-b) and contours of the six parameter model (c). Selected neighborhoods for two MTD curves when $p=0.1$ for pointwise percent selection determination (d). }
\label{fig:contours}
\end{figure}

When the working model is the logistic link function without an interaction term, the estimate $\bar{C}$ in (\ref{estmtd1}) is a decreasing line. Efficiency of the estimate of the MTD will be assessed by the pointwise average bias and percent selection. For $j=1,\ldots, m$, let $C_j$ be the estimated MTD curve (based on $n$ patients) and $C_{\rm true}$ be the true MTD curve. For every dose combination $(x,y) \in C_{\rm true}$, consider the relative minimum distance of $(x,y)$ to $C_j$

\begin{equation}
\label{distance}
d_{(x,y)}^{(j)}={\rm sign}(y'-y) \times {\rm min}_{(x^{\star},y^{\star}):(x^{\star},y^{\star}) \in C_j}  \left((x-x^{\star})^2+(y-y^{\star})^2)\right)^{1/2},
\end{equation}
where $y'$ is such that $(x,y') \in C_j$. The sign of $d_{(x,y)}^{(j)}$ depends on whether $(x,y)$ is above or below the estimated MTD curve $C_j$. The pointwise average bias is

\begin{equation}
\label{bias}
d_{(x,y)}= m^{-1} \sum_{j=1}^{m} d_{(x,y)}^{(j)}.
\end{equation}

Analogous to the percent selection for discrete dose combinations, we define the pointwise percent selection for continuous dose levels as follows. For each dose combination $(x,y) \in C_{\rm true}$, let $\Delta(x,y)$ be the Euclidean distance from the minimum dose combination to $(x,y)$. The percent selection is the fraction of trials $C_{j}, j=1,\ldots,m$ falling inside the circle with center $(x,y)$ and radius $p \Delta(x,y), 0< p <1$, where $p$ is a tolerance probability. This is a measure of percent of trials with MTD recommendation within $(100\times p)\%$ of the true MTD. Figure~\ref{fig:contours}(d) shows the variability of the radius of tolerance circles for various combinations on the MTD curve under two scenarios when the tolerance probability is $p=0.1$. Using the distance formula in (\ref{distance}), the pointwise percent selection with tolerance probability $p$ can be expressed as

\begin{equation}
\label{pselection}
m^{-1} \sum_{j=1}^{m} \mathbbm{1}\left(\left|d_{(x,y)}^{(j)}\right| \leq p\Delta(x,y)\right),
\end{equation}where $\mathbbm{1}(\cdot)$ is the indicator function.

\subsection{Results}
Table~\ref{tab:safety} shows that under scenarios 1 and 2, the average DLT rate across all simulated trials is close to the target $\theta = 0.33$ under the true and misspecified models when the working logistic model uses an interaction term.

\begin{table}[h]
\caption{Safety summary statistics under the four scenarios.}
\begin{tabular}{|c|c|c|c|c|}
\hline
Scenarios & Model & \makecell{Average \\ \% DLTs} & \makecell{\% Trials: \\ DLT rate \\ $> \theta + 0.05$} & \makecell{\% Trials: \\ DLT rate \\$> \theta + 0.1$} \\ \hline
\multirow{4}{*}{\begin{tabular}[c]{@{}c@{}}1 \\ $(\rho_{00}, \rho_{01}, \rho_{10}, \beta_3) = (0.01, 0.6, 0.6, 40)$ \end{tabular}} & Logistic & 33.57 & 7.80 & 0.70 \\ \cline{2-5} 
 & Probit & 31.97 & 2.90 & 0.05 \\ \cline{2-5} 
 & Log-Log & 31.98 & 4.35 & 0.10 \\ \cline{2-5} 
 & No-Interaction ($\beta_3 = 0$) & 36.07 & 18.30 & 0.95 \\ \hline
\multirow{4}{*}{\begin{tabular}[c]{@{}c@{}}2 \\ $(\rho_{00}, \rho_{01}, \rho_{10}, \beta_3) = (0.01, 0.2, 0.9, 100)$ \end{tabular}} & Logistic & 33.56 & 8.10 & 0.35 \\ \cline{2-5} 
 & Probit & 32.66 & 1.95 & 0.00 \\ \cline{2-5} 
 & Log-Log & 32.88 & 4.30 & 0.00 \\ \cline{2-5} 
 & No-Interaction ($\beta_3 = 0$) & 36.89 & 22.30 & 0.50 \\ \hline
\multirow{4}{*}{\begin{tabular}[c]{@{}c@{}}3 \\ $(\rho_{00}, \rho_{01}, \rho_{10}, \beta_3) = (0.001, 0.6, 0.01, 10)$ \end{tabular}} & Logistic & 26.01 & 0.05 & 0.00 \\ \cline{2-5} 
 & Probit & 25.93 & 0.00 & 0.00 \\ \cline{2-5} 
 & Log-Log & 25.84 & 0.00 & 0.00 \\ \cline{2-5} 
 & No-Interaction ($\beta_3 = 0$) & 30.17 & 0.45 & 0.00 \\ \hline
\multirow{2}{*}{\begin{tabular}[c]{@{}c@{}}4 \\ $(\alpha_1, \alpha_2, \alpha_3, \beta_1, \beta_2, \beta_3) = (0.5, 0.5, 2, 12, 5, 0.1)$ \end{tabular}} & Logistic & 24.25 & 0.30 & 0.00 \\ \cline{2-5} 
 & No-Interaction ($\beta_3 = 0$) & 29.98 & 2.45 & 0.10 \\ \hline
\end{tabular}
\label{tab:safety}
\end{table}


However, when omitting the interaction coefficient for the logistic working model under the true logistic model, this average DLT rate increases to 0.36 and 0.37 under these scenarios. Furthermore, the percent of trials with DLT rate exceeding $\theta+0.5$ can be at least 15\% higher when using a working model without an interaction term under scenario 1 and up to 20\% higher under scenario 2. A similar trend is observed under scenarios 3 and 4 but the trials are still safe with respect to the average DLT rate and percent of trials with a DLT rate exceeding $\theta+0.5$. In all cases, the percent of trials with an excessive rate of DLT (greater than $\theta+0.1$) is negligible. These results show that safety of the trial is compromised when the underlying synergism between the two drug is high and a model without an interaction term is used for dose finding.

\begin{figure}[h]
\caption{Pointwise average bias under the true and misspecified models under scenarios 1--4.  }
\centering
\subfloat[Scenario 1]{\includegraphics[width = 2.0in]{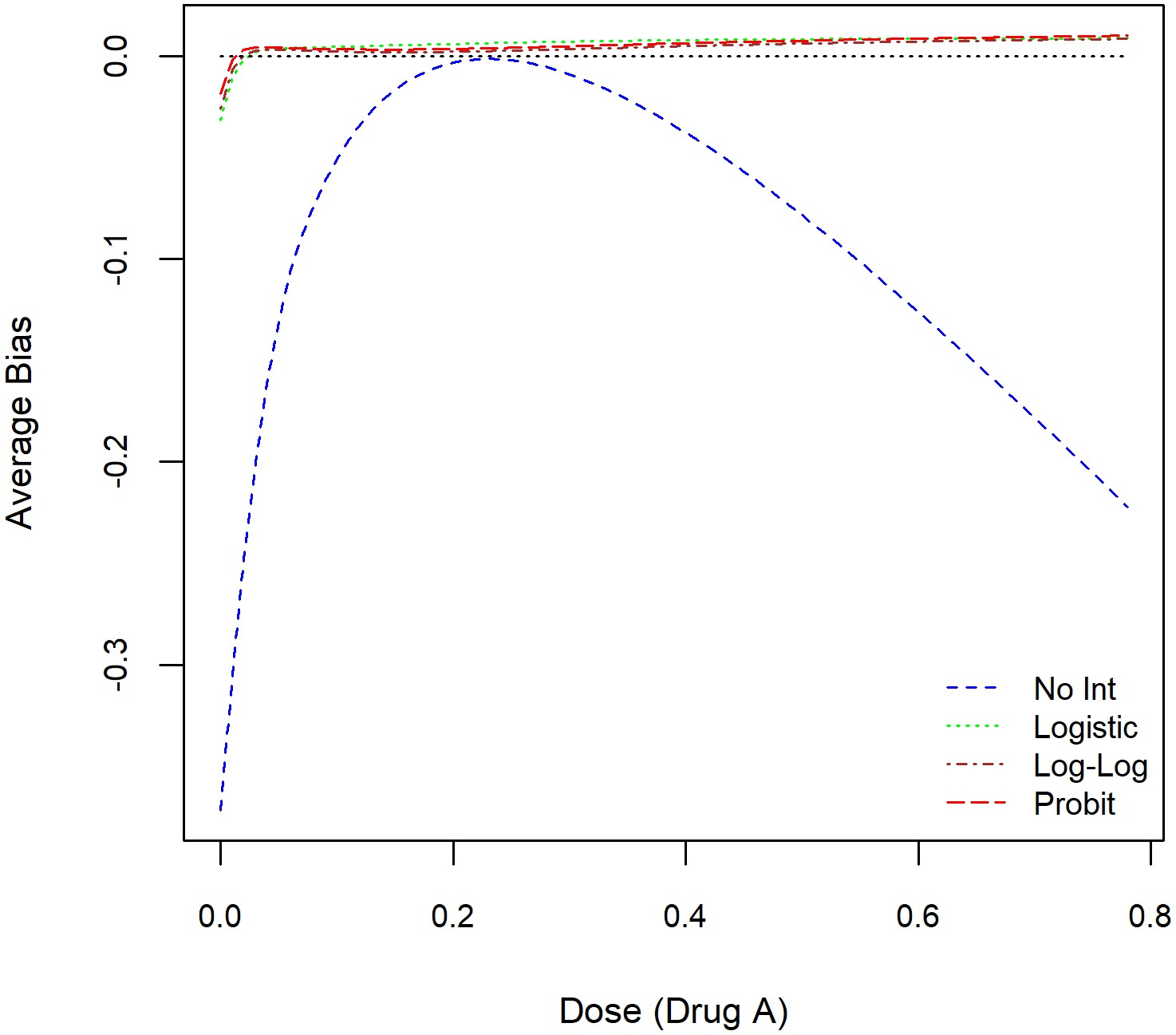}} 
\subfloat[Scenario 2]{\includegraphics[width = 2.0in]{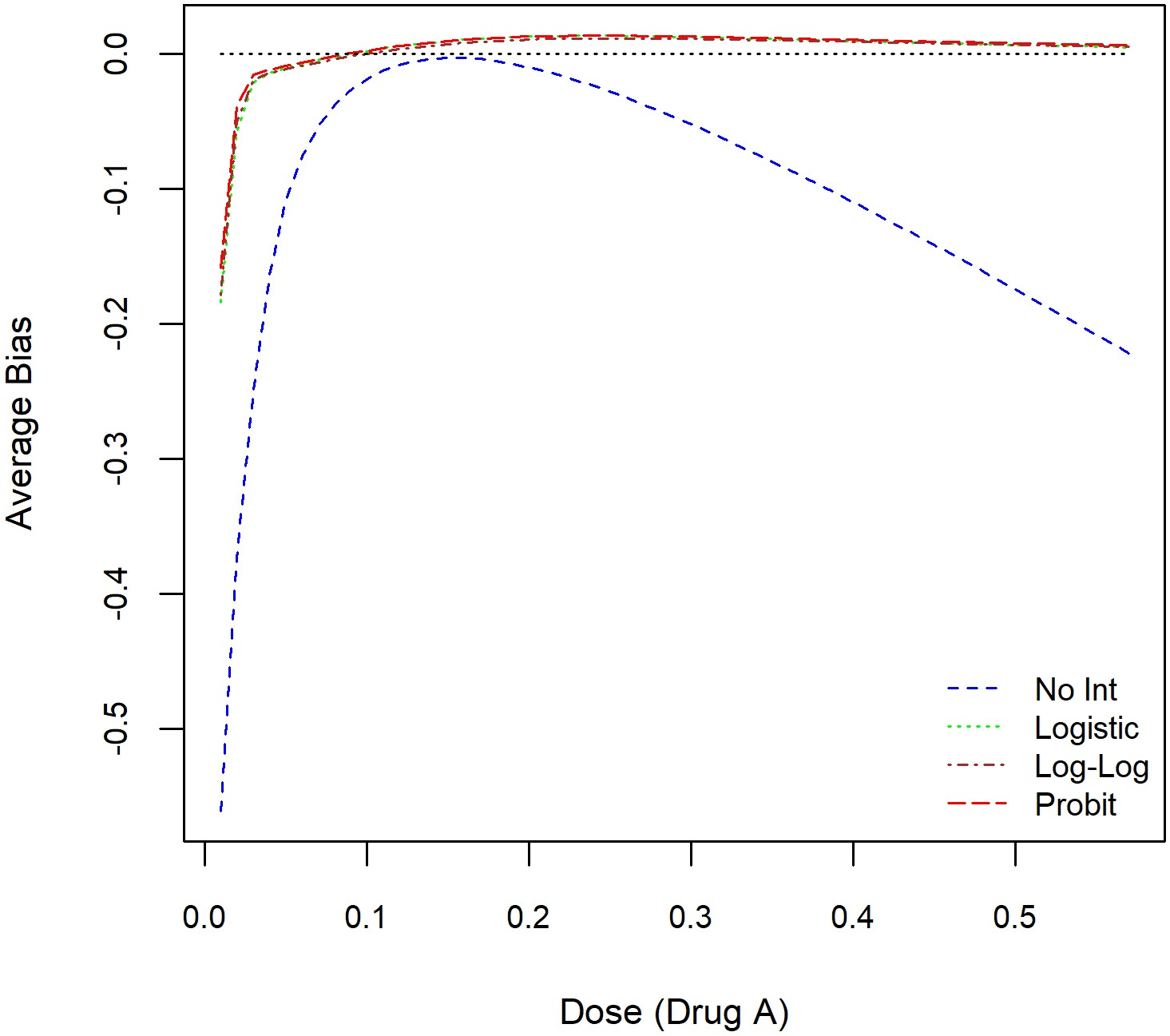}}\\
\subfloat[Scenario 3]{\includegraphics[width = 2.0in]{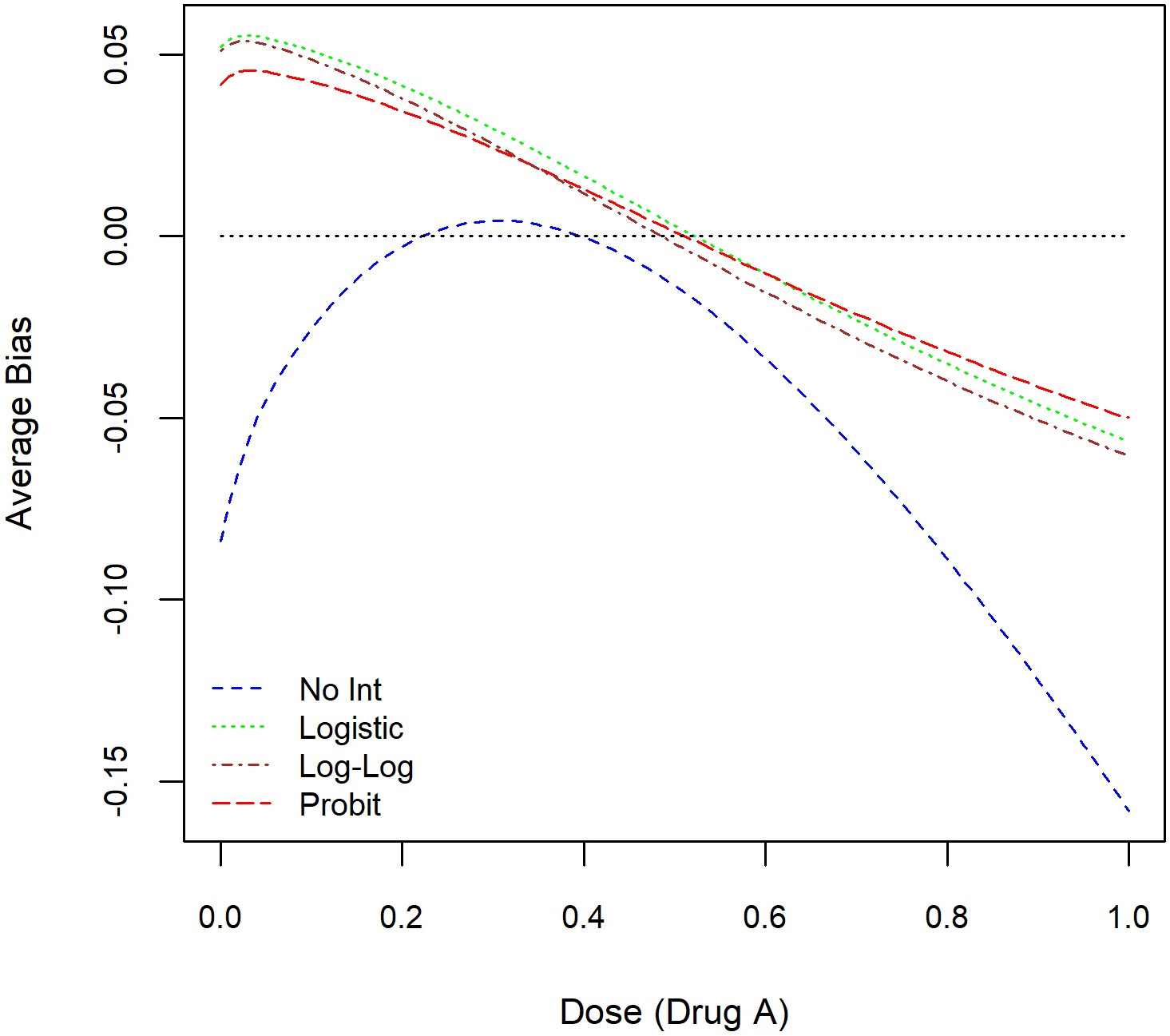}}
\subfloat[Scenario 4]{\includegraphics[width = 2.0in]{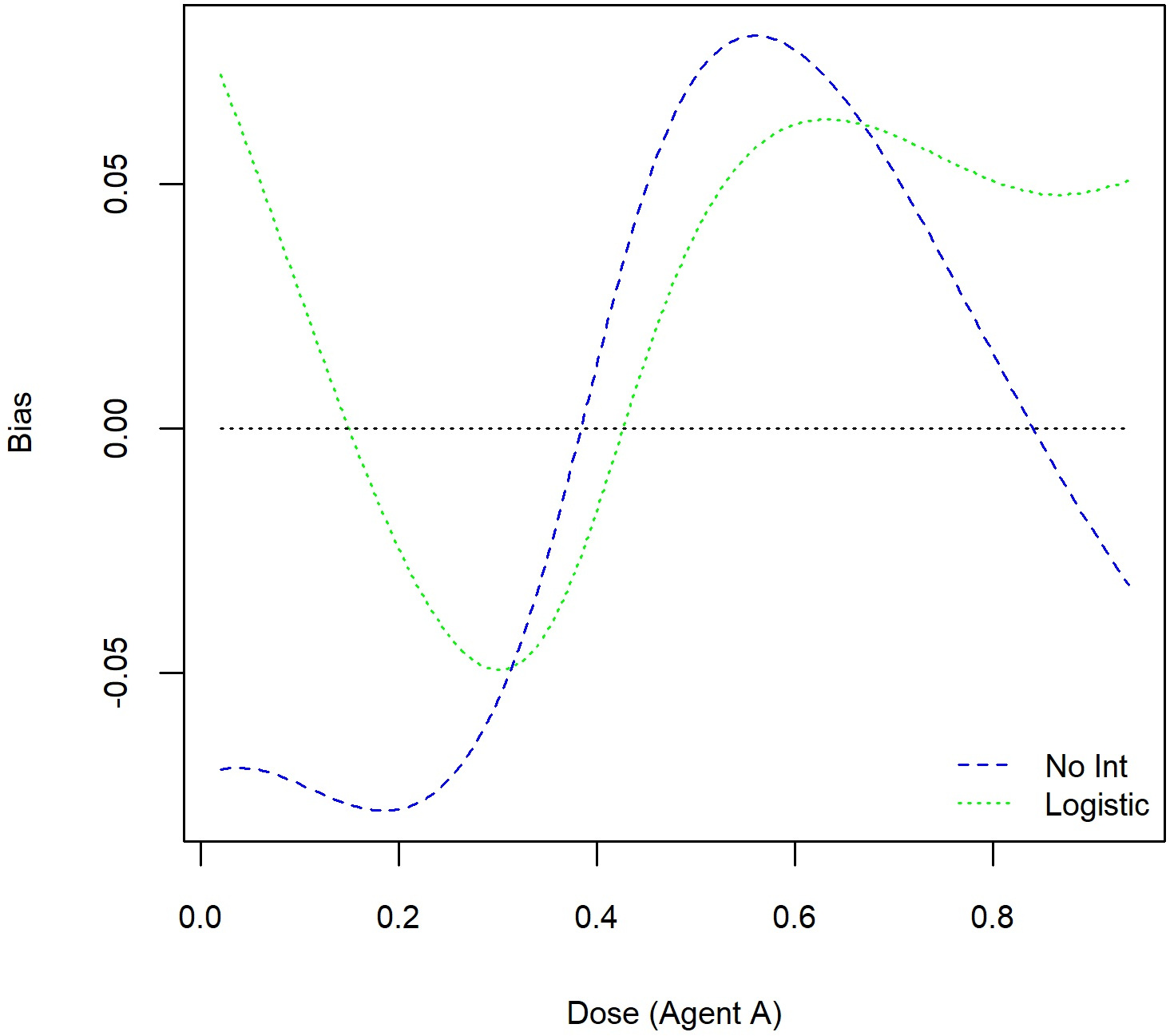}} 
\label{fig:bias}
\end{figure}

Figure~\ref{fig:scenarios} shows that the estimated MTD curves using the logistic, probit, complementary log-log working models are much closer to the true MTD curve relative to a logistic model without an interaction term, for the majority of dose combinations along the true MTD curve. This is also reflected in the pointwise average bias displayed in Figure~\ref{fig:bias}. Similar to the findings in \cite{tighiouart2014dose,tighiouart2017bayesian,2017bayesiandiniz, jimenez2019cancer}, the pointwise average bias is less than 5\% of the dose range except perhaps near the edges of the true MTD curve. However, omitting an interaction term leads to a substantial increase in this average bias by as much as 20\% of the dose range under scenario 1 for standardized doses of drug $A$ around 0.1 or 0.7 and around 0.05 and 0.5 under scenario 2. Under scenario 3, when using the logistic model without the interaction coefficient, the absolute pointwise average bias varies between 10\% and 15\% of the dose range when the dose level of drug $A$ varies between 0.8 and 1.0 compared to 2.5\% and 5\% of the dose range for models that include an interaction term. For scenario 4, the pointwise average bias for the logistic model with and without an interaction term are less than 7.5\% of the dose range for almost all dose combinations. 

\begin{figure}[h]
\centering
\caption{Pointwise percent selection under the true and misspecified models under scenarios 1--4 for a tolerance probability $p=0.1.$ }
\subfloat[Scenario 1]{\includegraphics[width = 2.0in]{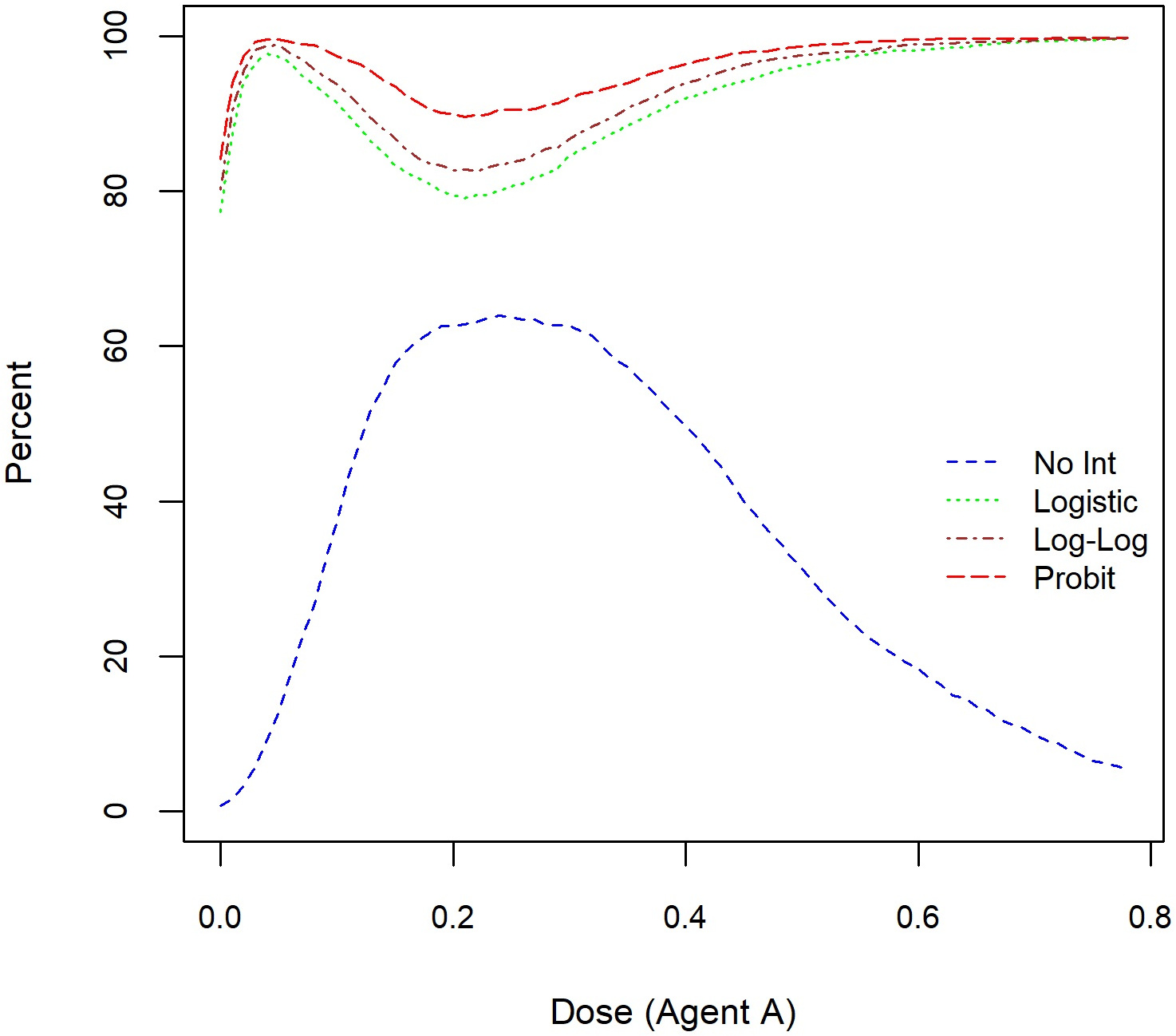}} 
\subfloat[Scenario 2]{\includegraphics[width = 2.0in]{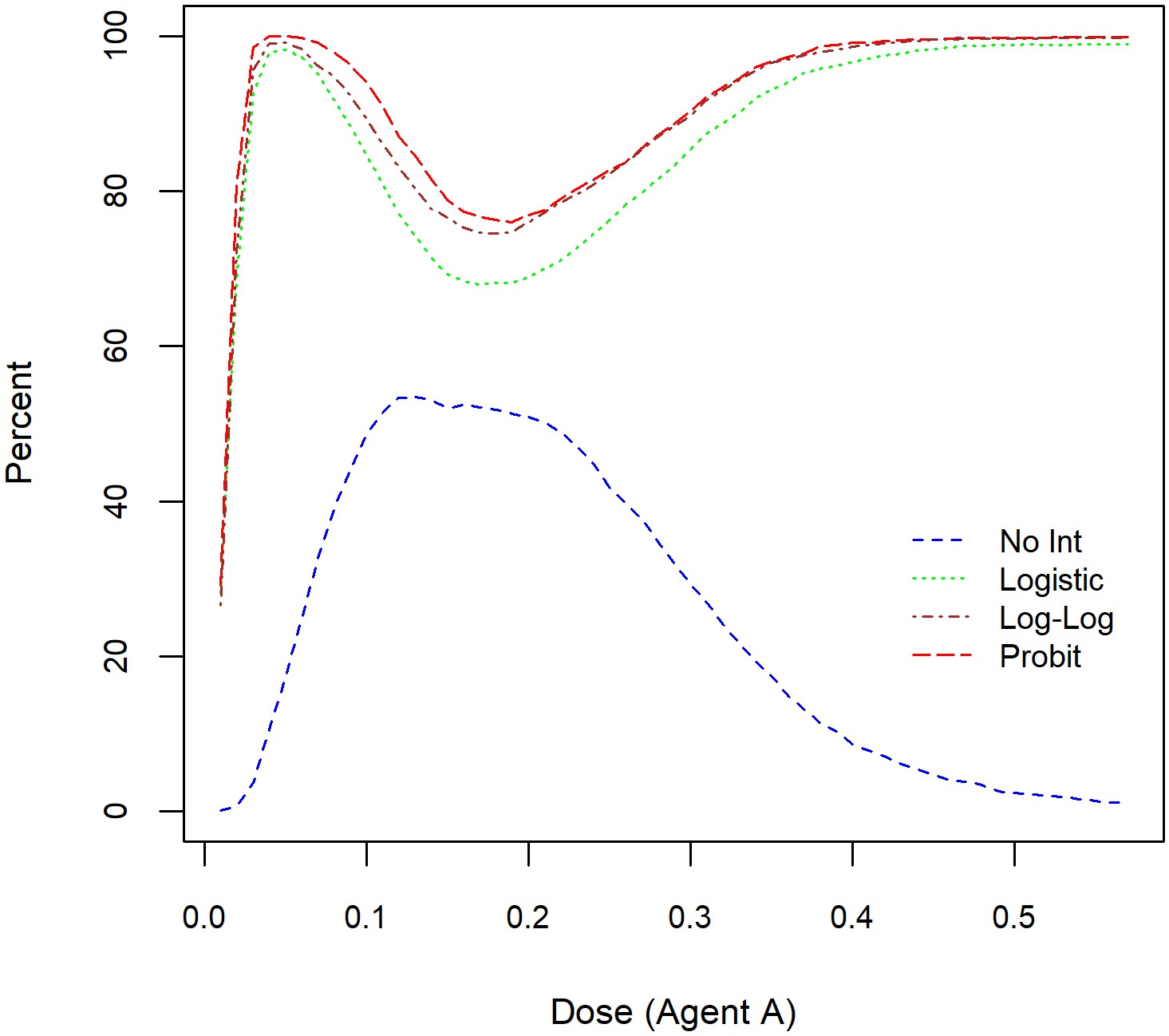}}\\
\subfloat[Scenario 3]{\includegraphics[width = 2.0in]{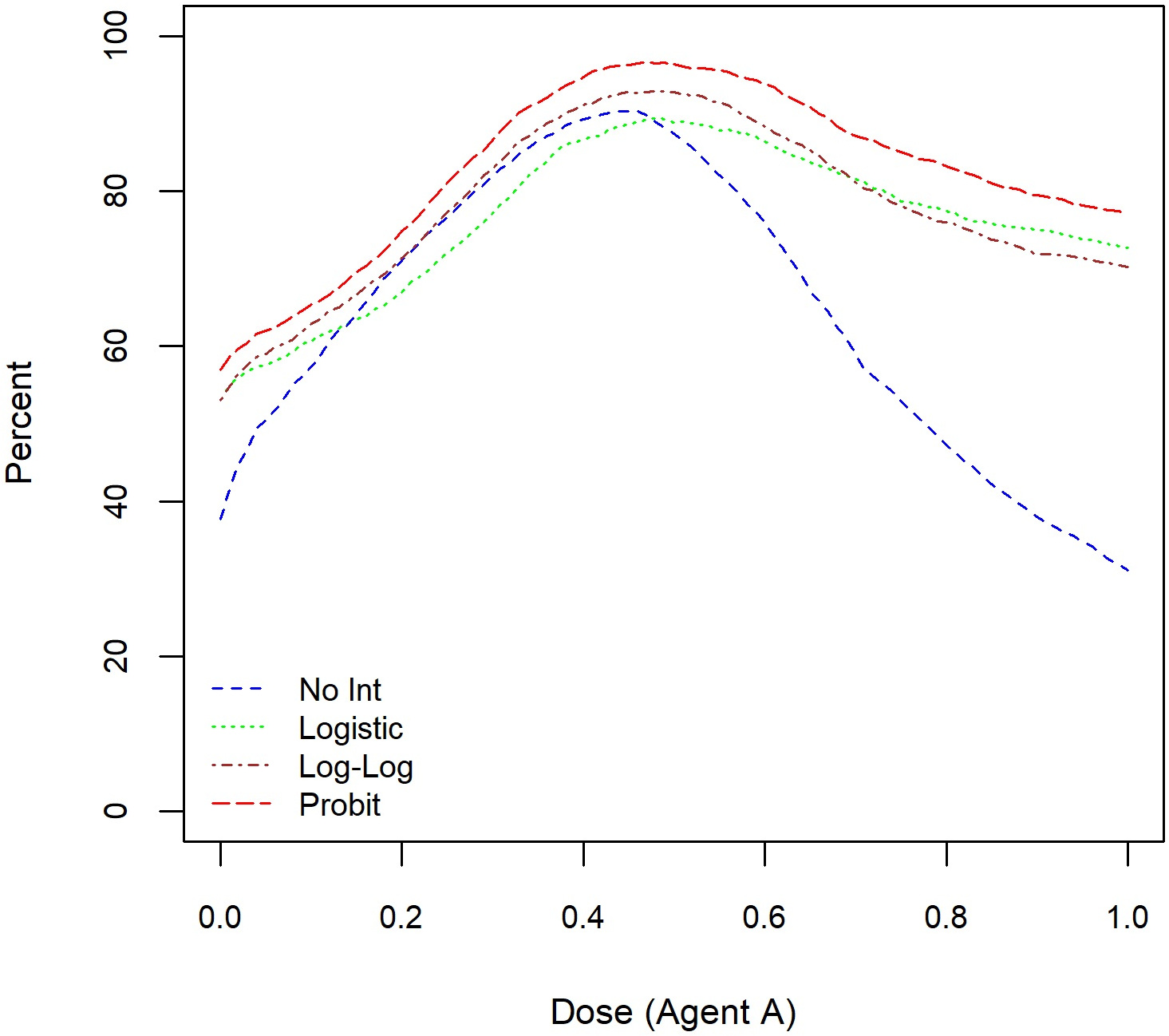}}
\subfloat[Scenario 4]{\includegraphics[width = 2.0in]{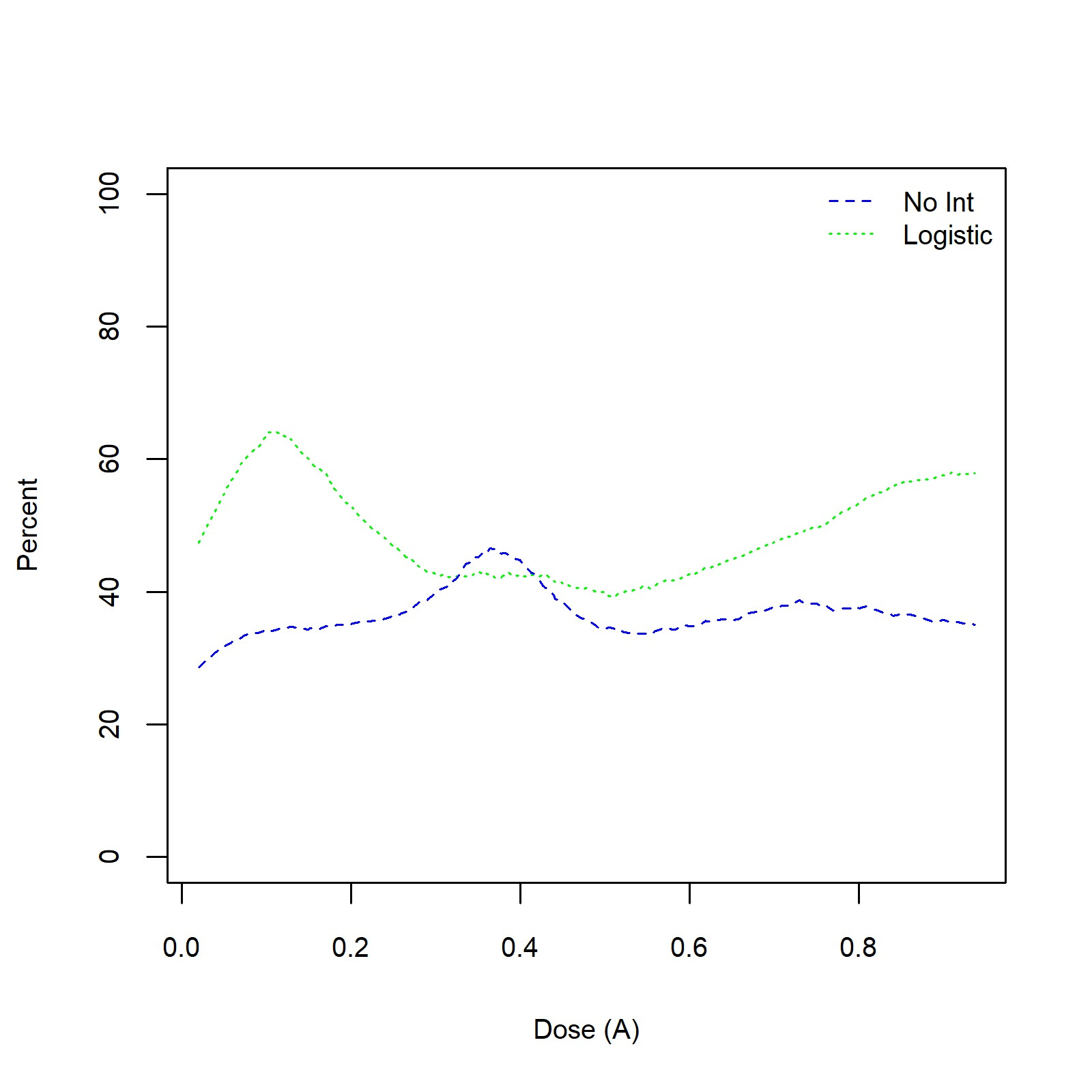}} 
\label{fig:pselection}
\end{figure}

In Figure \ref{fig:pselection}, the pointwise percent selection using a tolerance probability $p=0.1$ is very poor when the two drugs are highly synergistic $(\beta_3=40, 100)$ under scenarios 1 and 2 for the logistic model without an interaction coefficient. Under scenario 1, this percent selection varies between 80\% and 100\% when using all three models with interaction terms - logistic, probit, and complementary log-log - compared to a 0\% to 60\% range when using a logistic model with $\beta_3=0$. 

A similar poor performance is observed under scenario 2 and to a lesser extent for scenario 3 where there is an overlap between the models when the dose level of drug $A$ is less than 0.5. For dose combinations above 0.5, this pointwise percent selection drops considerably relative to models that include the interaction term. Under scenario 4 where the true dose-toxicity relationship is the six-parameter model, the pointwise percent selection using the logistic model with an interaction term is still higher relative to the model without an interaction parameter for most dose combinations and the extent of difference can be as high as 30\% near dose level 0.15 of drug $A$. Therefore, we conclude that the efficiency of the estimate of the MTD curve as measured by the pointwise bias and percent selection is greatly reduced when omitting an interaction coefficient in the model. 

\begin{figure}[h]
\centering
\caption{Two-dimensional density plots of the last doses from $m=2000$ trial replicates under scenarios 1--4.}
\subfloat[Scenario 1- With Interaction]{\includegraphics[width = 1.8in]{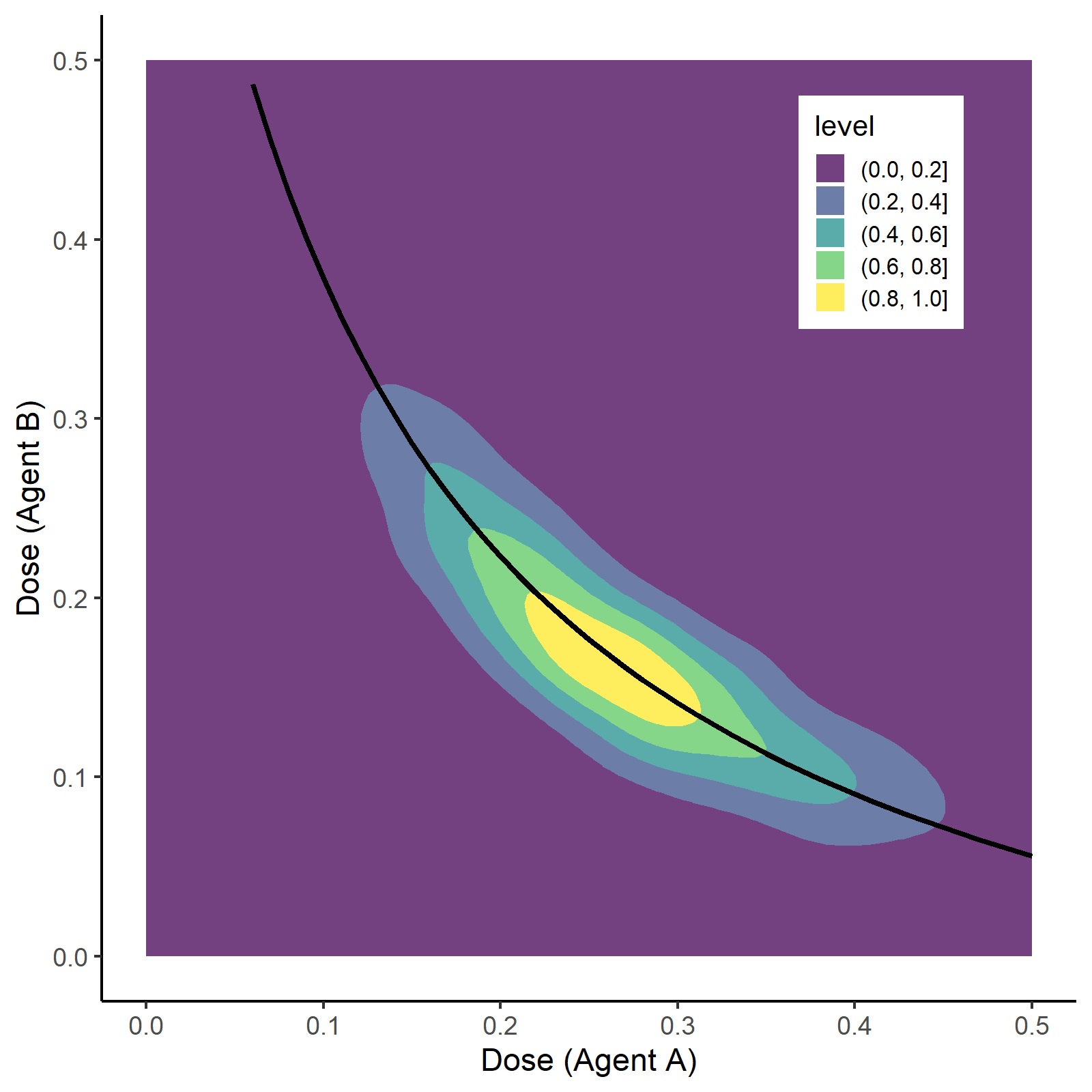}} 
\subfloat[Scenario 1- Without Interaction]{\includegraphics[width = 1.8in]{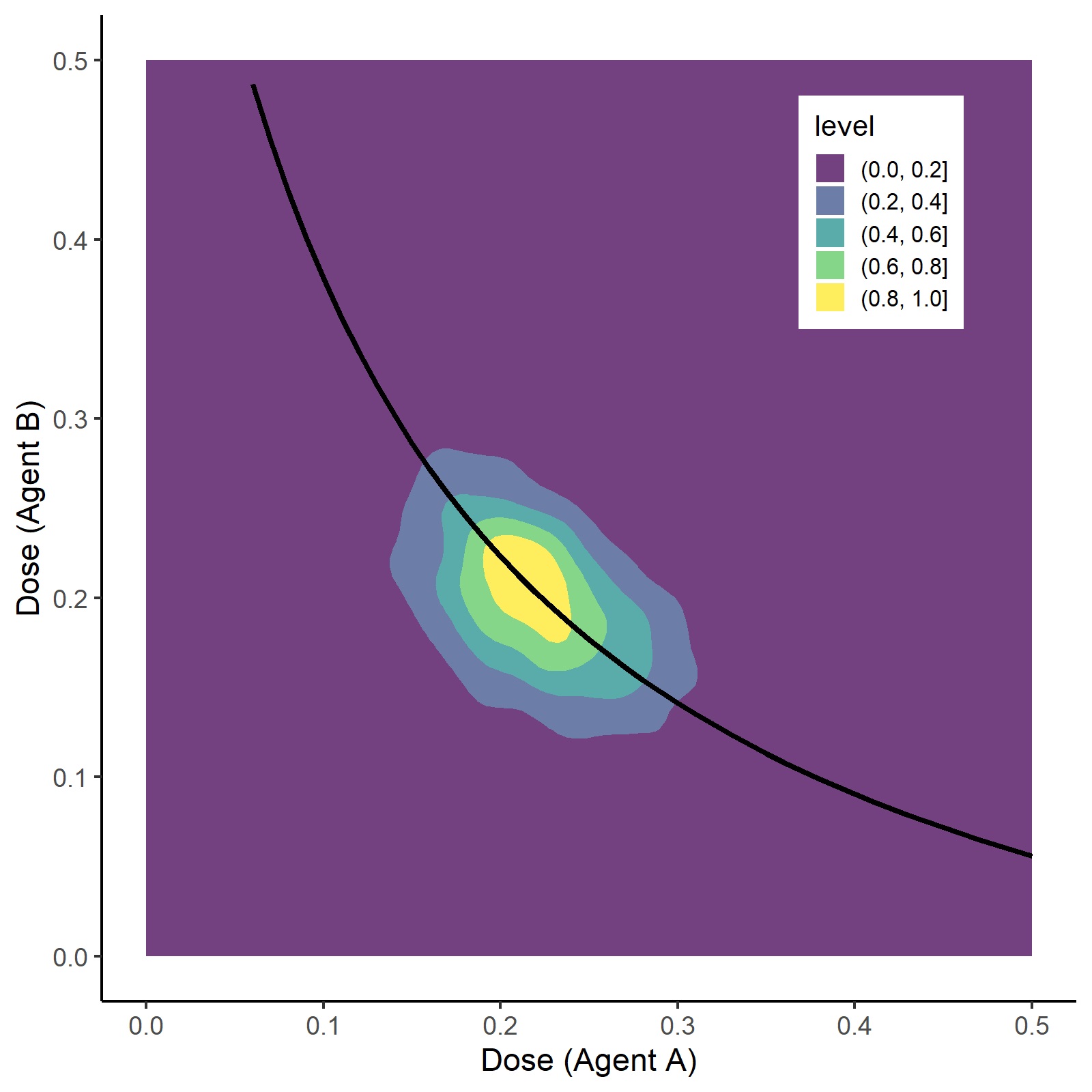}}\\
\subfloat[Scenario 2- With Interaction]{\includegraphics[width = 1.8in]{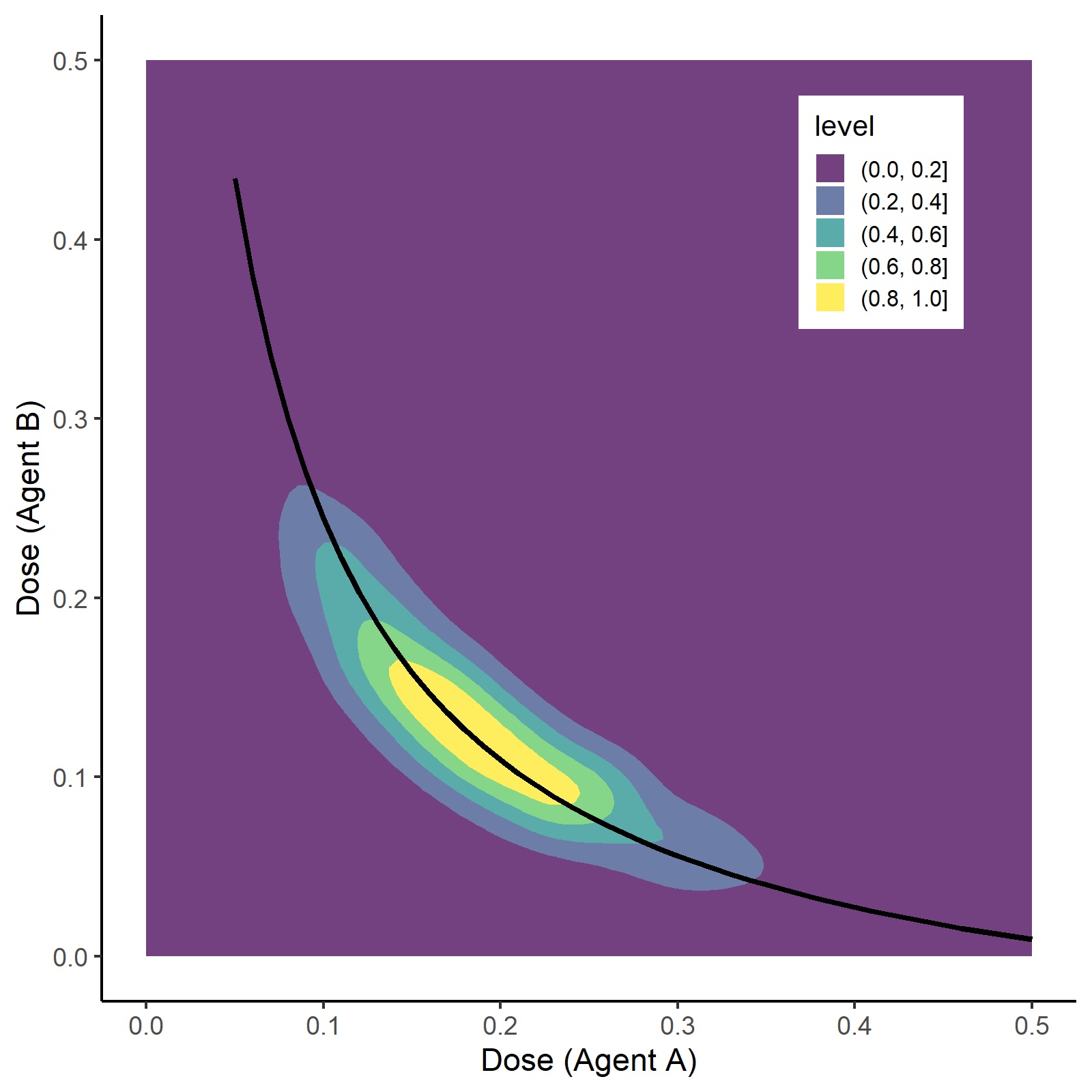}}
\subfloat[Scenario 2- Without Interaction]{\includegraphics[width = 1.8in]{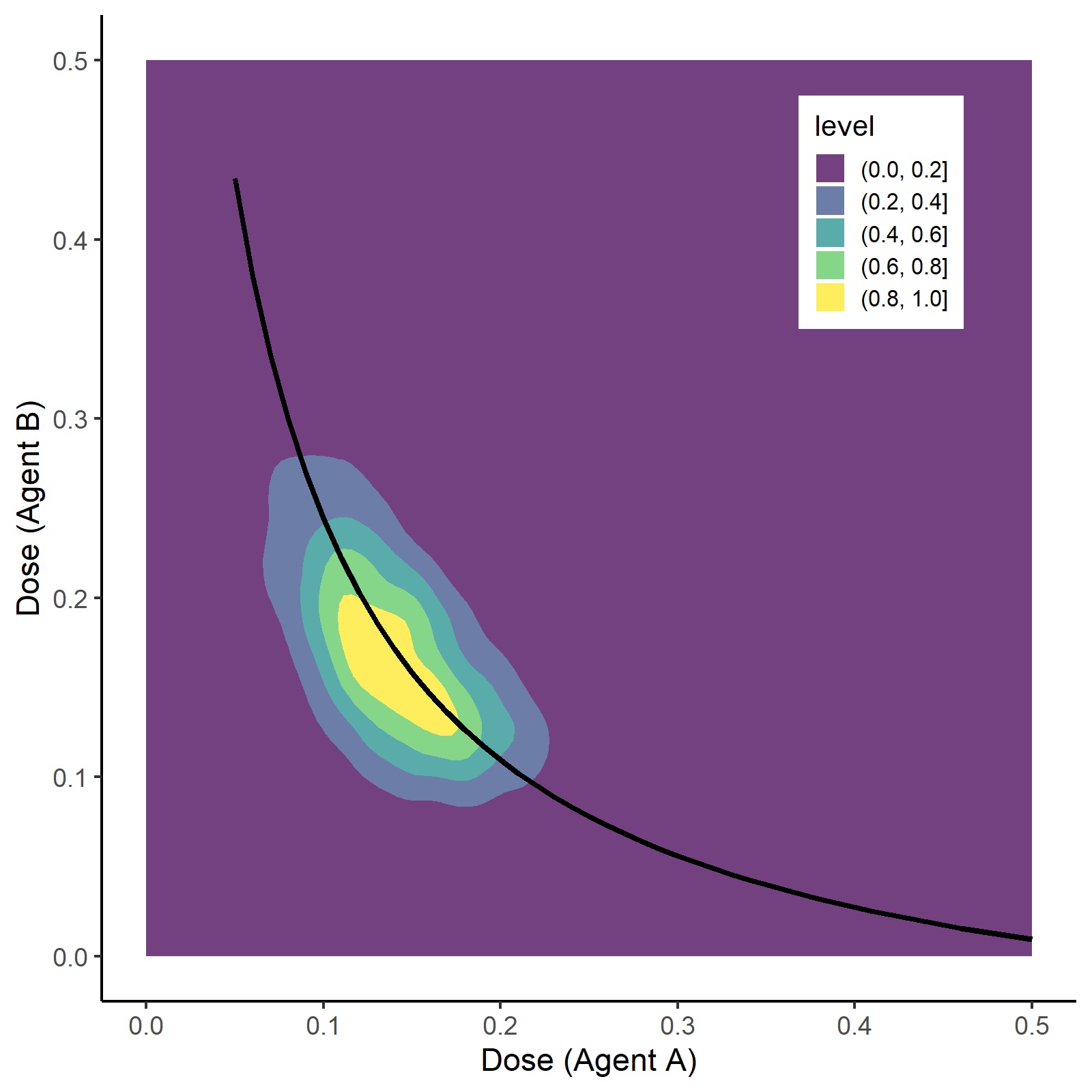}} \\
\subfloat[Scenario 3- With Interaction]{\includegraphics[width = 1.8in]{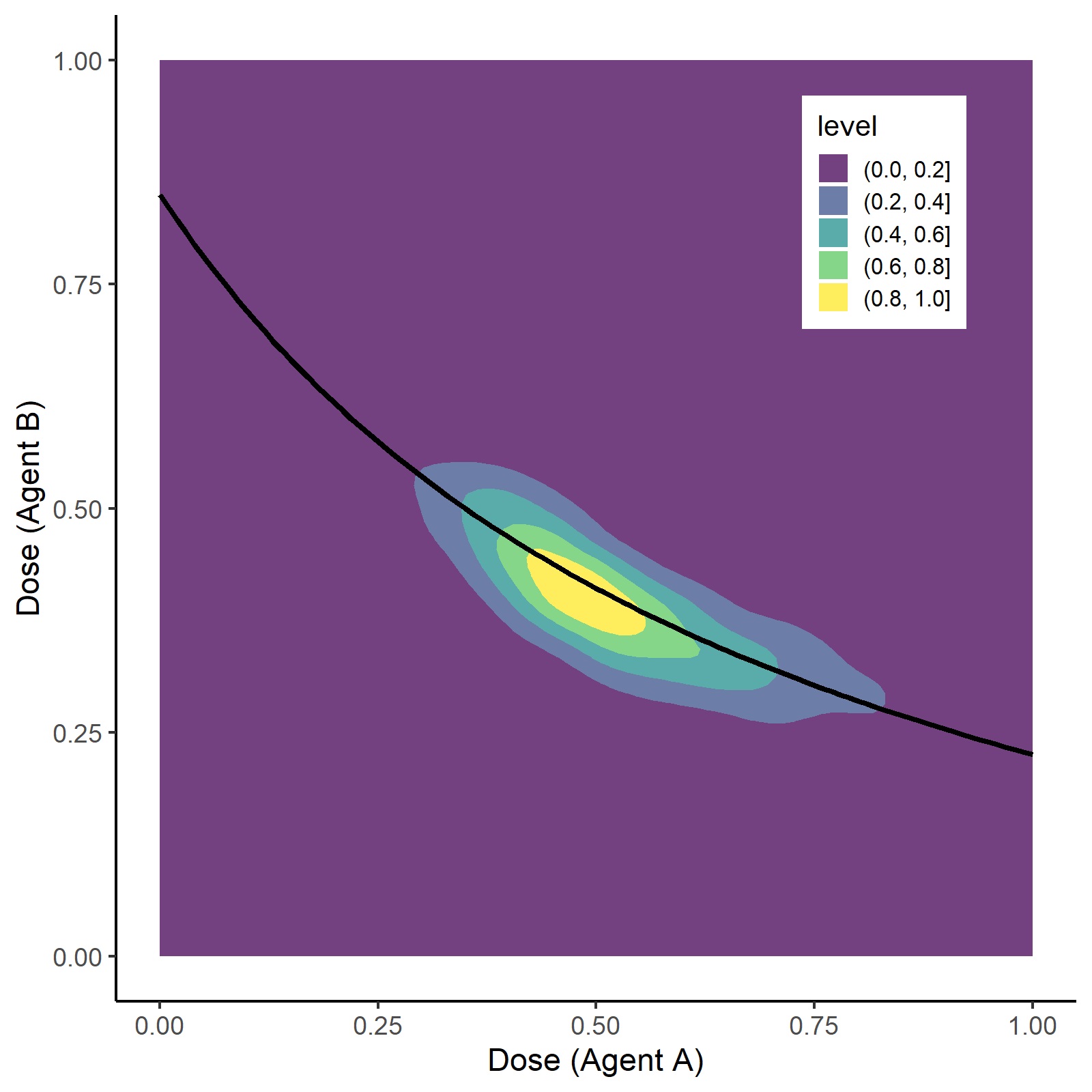}} 
\subfloat[Scenario 3- Without Interaction]{\includegraphics[width = 1.8in]{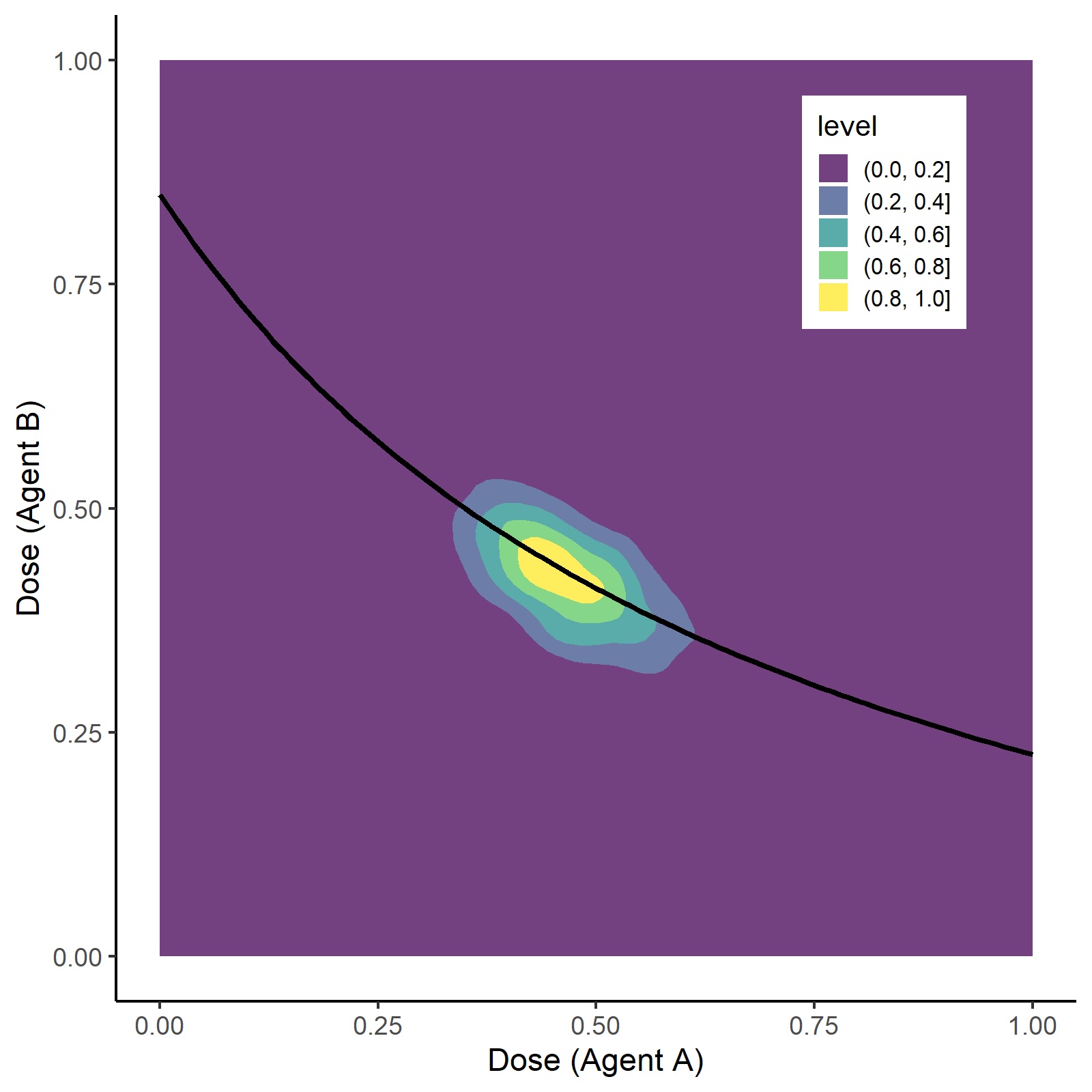}}\\
\subfloat[Scenario 4- With Interaction]{\includegraphics[width = 1.8in]{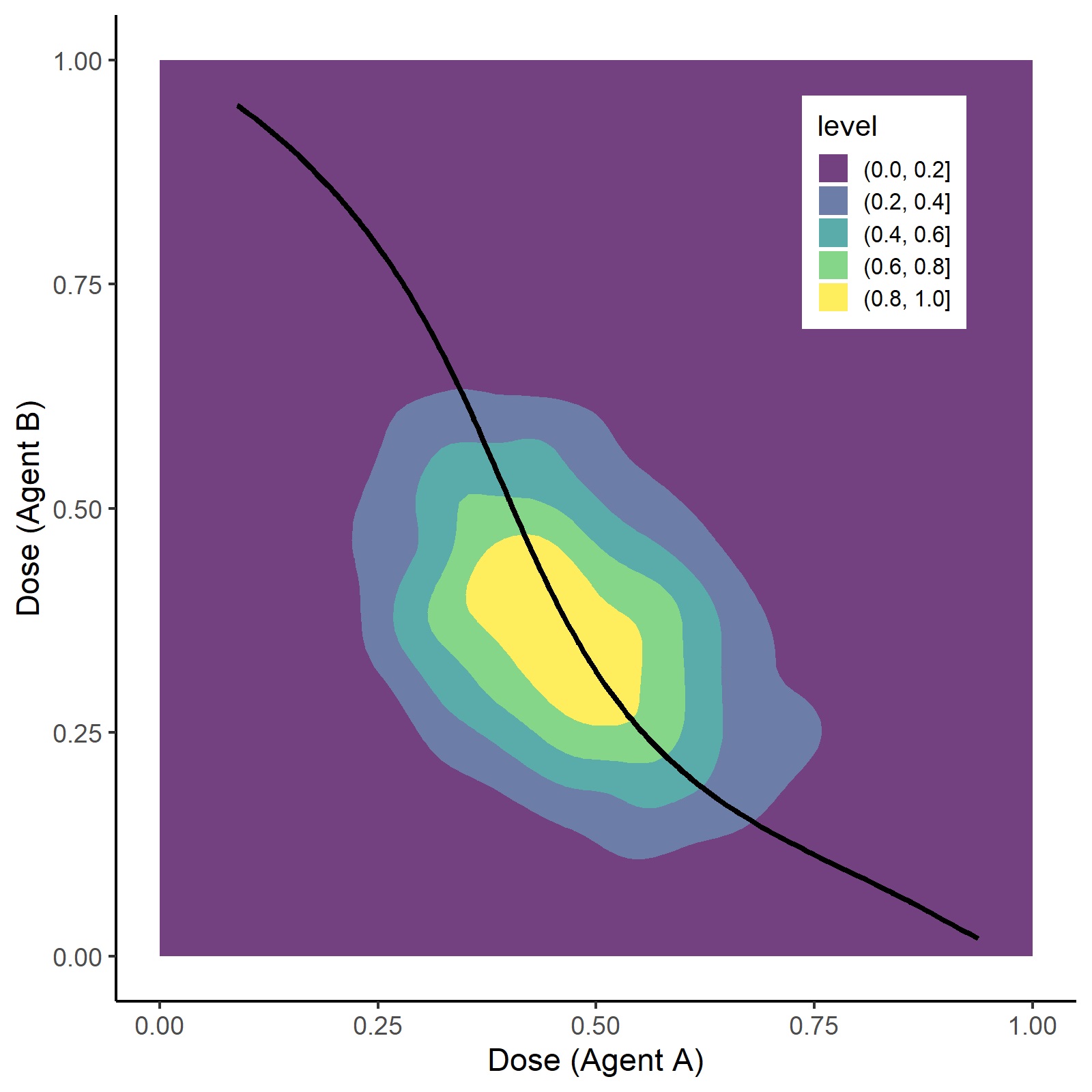}}
\subfloat[Scenario 4- Without Interaction]{\includegraphics[width = 1.8in]{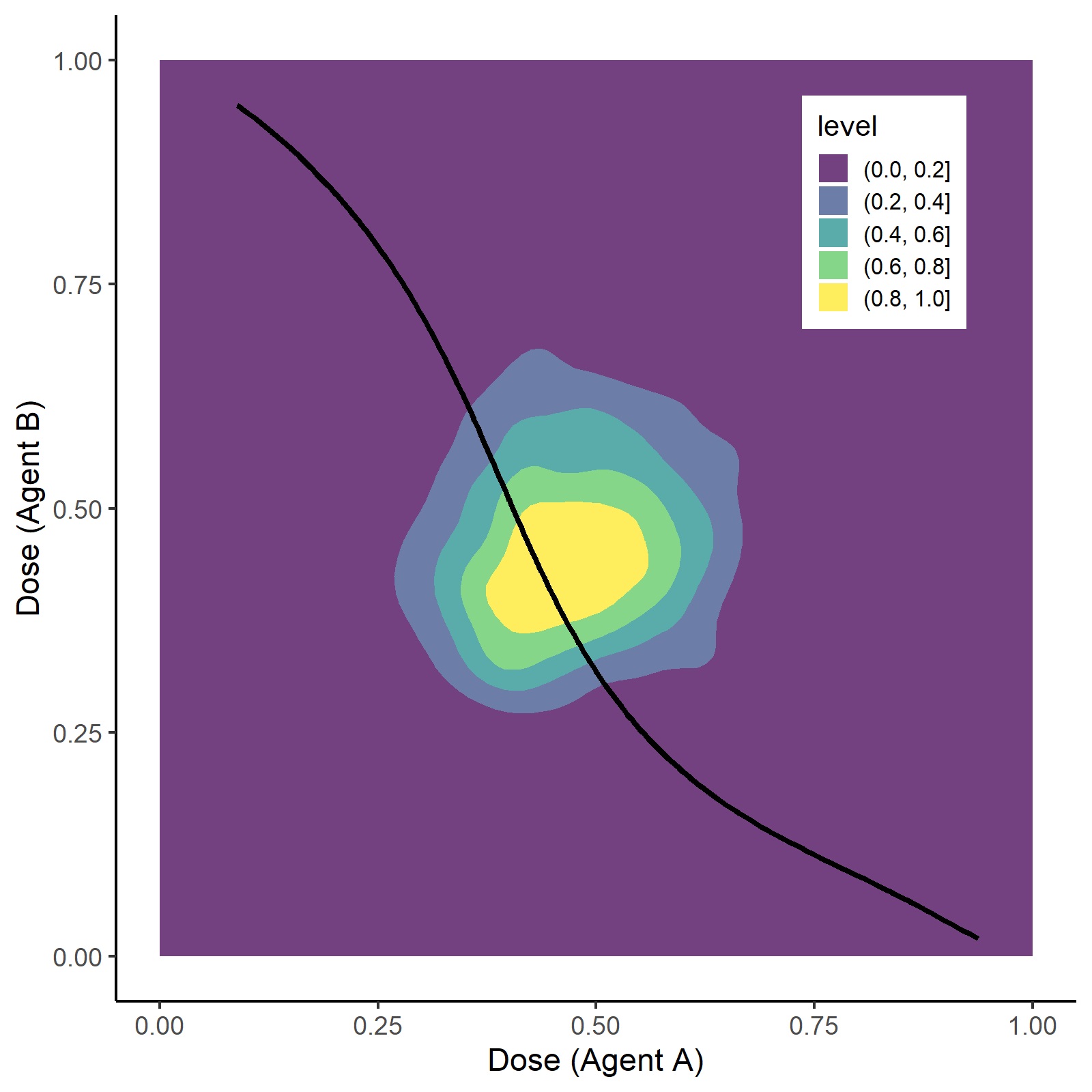}}
\label{fig:density-lastdose}
\end{figure}

For trial designs that recommend a single MTD, usually taken as the last dose combination given to the last patient in the trial or the dose that would have been given to the $(n+1)$th hypothetical patient in the trial \cite{Braun2010,wages2011continual}, Figure~\ref{fig:density-lastdose} shows the two-dimensional density plots of the last dose combination from all $m=2000$ simulated trials under each scenario, using the logistic model with and without an interaction term. These plots show that using a model with an interaction term results in a much wider coverage area along the MTD curve relative to a model without an interaction coefficient. For example, under scenario 1, starting with density level 0.2, the coverage area for the logistic model with an interaction term is $[0.1, 0.45] \times [0.08, 0.32]$ compared to the dose combination range of $[0.15, 0.32] \times [0.14, 0.28]$ when omitting the interaction coefficient. The later model seems to recommend a dose that is concentrated near the middle of the curve with high probability. Therefore, if dose combinations away from the middle of the true MTD curve have high probability of efficacy relative to doses in the middle of the curve, then using models with no interaction term will more likely result in a failed phase II efficacy trial compared to models that include an interaction coefficient.

\section{Conclusions}

The use of drug combinations in cancer treatment allows targeting of different signaling pathways simultaneously, may reduce treatment resistance to therapy, clonal heterogeneity, and suppress cellular mechanisms associated with adaptive resistance. When several dose levels of these combinations are explored in early phase trials, more than one maximum tolerated doses can exist with possibly different levels of efficacy. This multiplicity problem is further amplified by the inherent small sample size of these trials, and this has led many authors to use parsimonious models for dose finding, especially in the absence of prior information about each drug when used individually. In fact, several authors recommended dropping the interaction term modeling synergism between the two drugs without affecting the performance of the design. They argued that precise estimation of the interaction coefficient is not possible due to the small sample size and omitting this term has negligible impact on the percent of dose recommendations at the end of the trial. These claims have been assessed for discrete dose combinations under one particular logistic model and a design that recommends one MTD by \cite{Mozgunov2021} and were found to be consistent with the previous recommendations.

We evaluated these assertions in the setting of drug combinations with continuous dose levels with a design that recommends an MTD curve at the end of the trial. Contrary to the findings of the previous authors, we found that when the two drugs are highly synergistic, not including an interaction term can result in a significantly higher percent of trials with a DLT rate exceeding $\theta+0.05$ and hence compromises the safety of the trial. In addition, efficiency of the estimated MTD curve as assessed by the pointwise average bias and percent selection is greatly reduced relative to a design that includes the interaction coefficient. We note that these findings were obtained when the true dose-toxicity model belongs to the family $F(\beta_{0}+\beta_{1}x+\beta_{2}y+\beta_{3}xy)$ for different link functions and the six-parameter model $(1+\alpha_{1}x^{ \beta_1}+\alpha_{2}y^{\beta_2}+\alpha_{3}(x^{\beta_1}y^{\beta_2})^{\beta_3})^{-1},$ and the working model uses the logistic link function. In addition, the algorithm proposed by \cite{tighiouart2017bayesian} was used for dose escalation and MTD curve estimation. Therefore, these conclusions may not apply to other forms of dose-toxicity models and dose allocation algorithms. We further showed that even for some designs that recommend a single MTD at the end of the trial, omitting an interaction term may result in a dose recommendation at the end of the trial with low probability of efficacy. In practice, for either discrete or continuous dose levels, we recommend that the statistician derives the operating characteristics in collaboration with the clinician using a model with and without an interaction term under a large class of scenarios that assume weak and very strong synergism between the two drugs.

\subsection*{Acknowledgments}

This work is partially funded by NIH the National Center for Advancing Translational Sciences (NCATS) UCLA CTSI (UL1 TR001881-01), NCI grant P01 CA233452-02, and U01 grant CA232859-01.

\subsection*{Conflicts of Interest}

Jos\'e L. Jim\'enez is employed by Novartis Pharma A.G. who provided support in the form of salary for the author, but did not have any additional role in the preparation of the manuscript. Also, the views expressed in this publication are those of the authors and should not be attributed to any of the funding institutions or organisations to which the authors are affiliated.

\bibliographystyle{plainnat}
\bibliography{references2022}

\end{document}